\documentclass[oldversion]{aa}

\usepackage{graphicx}
\usepackage{txfonts}
\usepackage[round]{natbib}
\usepackage{amssymb}
\usepackage[figuresright]{rotating}

\bibpunct{(}{)}{;}{a}{}{,}
\def\ion#1#2{{\rm #1}\,{\sc #2}}

\begin{document}
\title{A new comprehensive set of elemental abundances in DLAs \\ 
III. Star formation histories
\thanks{Based on UVES observations made with the European Southern Observatory 
VLT/Kueyen telescope, Paranal, Chile, collected during the programme ID 
No.~70.B--0258(A).}}

%\subtitle{}

\author{
       M. Dessauges-Zavadsky\inst{1},
       F. Calura\inst{2},
       J. X. Prochaska\inst{3},
       S. D'Odorico\inst{4},	   
       \and
       F. Matteucci\inst{5,2}
        }

%\offprints{M. Dessauges-Zavadsky, \\ \email{miroslava.dessauges@obs.unige.ch}}
\offprints{miroslava.dessauges@obs.unige.ch}

\institute{
	  Observatoire de Gen\`eve, Universit\'e de Gen\`eve, 51 Ch. des 
	  Maillettes, 1290 Sauverny, Switzerland %\\
	  %\email{miroslava.dessauges@obs.unige.ch}
	  \and
	  INAF - Osservatorio Astronomico di Trieste, Via G. B. Tiepolo 11, 
	  34131 Trieste, Italy %\\
	  %\email{fcalura@ts.astro.it}
	  \and
	  UCO/Lick Observatory, University of California at Santa Cruz, Santa 
	  Cruz, CA 95064, USA %\\ 
	  %\email{xavier@ucolick.org}
	  \and
	  European Southern Observatory, Karl-Schwarzschildstr. 2, 85748 
	  Garching bei M\" unchen, Germany %\\  
	 %\email{sdodoric@eso.org}
	  \and
	  Dipartimento di Astronomia, Universit\'a di Trieste, via G.B. 
	  Tiepolo 11, 34131 Trieste, Italy %\\
	  %\email{matteucci@ts.astro.it}
           }

\date{Received; accepted}

\authorrunning{M. Dessauges-Zavadsky et al.}

\titlerunning{Star formation histories}

\abstract{
We obtained comprehensive sets of elemental abundances for eleven damped 
Ly$\alpha$ systems (DLAs) at $z_{\rm DLA} = 1.7-2.5$. For nine of them, we 
accurately constrained their intrinsic abundance patterns accounting for dust 
depletion and ionization effects. In Paper~I of this series,
we showed for three DLA galaxies that we can derive their star formation 
histories and ages from a detailed comparison of their intrinsic abundance 
patterns with chemical evolution models. We determine in this paper the star 
formation properties of six additional DLA galaxies. The derived results 
confirm that no single star formation history explains the diverse sets of 
abundance patterns in DLAs. We demonstrate that the various star formation 
histories reproducing the DLA abundance patterns are typical of local 
irregular, dwarf starburst and quiescent spiral galaxies. Independent of the 
star formation history, the DLAs have a common characteristic of being weak 
star forming galaxies; models with high star formation efficiencies are ruled 
out. The distribution of the DLA star formation histories shows a trend of 
finding more galaxies with a star formation history typical of dwarf irregulars 
with a bursting star formation toward high redshifts, $z > 2$. Only two DLA 
galaxies (each at $z < 2$) in our sample of nine objects have a star formation 
history typical of spiral galaxies. Since DLAs sample the broad distribution of 
galaxies at high redshift, this trend indicates that young and less evolved 
proto-galactic structures with low masses and low star formation rates are more 
common toward higher redshifts. This is further supported by the star formation 
rate and age distributions. Indeed, all the derived DLA star formation rates 
per unit area are moderate or low, with values between $-3.2 < \log {\rm SFR} 
< -1.1$ M$_{\odot}$~yr$^{-1}$~kpc$^{-2}$. 
%namely $10-1000$ lower than in emission-selected galaxies known to be biased 
%toward luminous and strong star formation galaxies. 
The DLA abundance patterns also require a large spread in ages ranging from 20 
Myr up to 3 Gyr. Enhanced $\alpha$ over iron-peak ratios are associated with 
young objects having undergone a recent burst of star formation, while solar 
$\alpha$ over iron-peak ratios are associated with old objects undergoing an 
inefficient continuous star formation. The oldest DLA in our sample is observed 
at $z_{\rm DLA} = 1.864$ with an age estimated to more than 3 Gyr; it nicely 
indicates that galaxies were already forming at $z_f\gtrsim 10$. But, most of 
the DLAs show ages much younger than that of the Universe at the epoch of 
observation. Young galaxies thus seem to populate the high redshift Universe at 
$z > 2$, suggesting relatively low redshifts of formation ($z\sim 3$) for most 
high-redshift galaxies. The large dispersion in star formation history and age
indicates that the DLAs are drawn from a diverse population of galaxies. The 
DLA star formation properties are compared with those of other high-redshift 
galaxies identified in deep imaging surveys with the aim of obtaining a global 
picture of high-redshift objects.
%JXP -- I did a lot of edits in the abstract, so re-read and re-edit as
%  you desire.  I would also suggest adding a sentence which is somewhat
%  obvious along the lines of: ``The large dispersion in age, SFH and XXX
%  indicate the DLA are drawn from a diverse population of galaxies.''

}

\maketitle

%
%________________________________________________________________

\section{Introduction}

Quasar (QSO) absorption line systems provide an alternative means to study the
high-redshift galaxy properties other than the traditional stellar luminosity
selected galaxy samples \citep[][ and reference therein]{steidel03}. Indeed, 
they are expected to select galaxies based on gas cross-section, rather than 
luminosity, and hence give a complementary view of the galaxy population to 
that obtained from emission-selected galaxies. The damped Ly$\alpha$ systems 
(DLAs) with \ion{H}{i} column densities higher than $2\times 10^{20}$ cm$^{-2}$ 
are particularly interesting, since they contain the majority of the neutral 
gas at high redshifts and may provide the fuel reservoirs for subsequent star 
formation in galaxies \citep[e.g.][]{wolfe86,prochaska05}. 

The connection between the population of DLA absorbers and galaxy populations
identified at high redshifts in deep imaging studies remains, however, a key 
topic in extragalactic astronomy today. While an evolutionary link between the 
emission-selected galaxies at $z=2-3$ and the bulge component of the galaxy 
population today is strongly suggested \citep{steidel96,adelberger04,steidel05}, 
an analogous connection for DLAs is still a matter of debate, because of the 
difficult detection of DLA galaxy counterparts \citep{moller04}. Indeed, the 
complete picture relating the gas cross-section, ionization and chemical 
abundances of the interstellar media and gaseous halos surrounding galaxies to
the emission properties of the associated galaxy remains incomplete.

DLA galaxy counterparts are difficult to identify, because they typically lie 
at angular separations of the order of one arcsec from the quasar line of 
sight, and are therefore swamped by the light from the quasar. At high redshift 
($z > 2$), despite large observational efforts, emission, in general from 
Ly$\alpha$, from seven DLA counterparts was detected and spectroscopically 
confirmed, but only four are not associated DLAs with the QSO, $z_{\rm DLA} \ll
z_{\rm QSO}$ \citep[see a summary by][]{weatherley05}. At lower redshift 
($z < 1$), the DLA galaxy counterparts are more readily observed. Deep-imaging 
and follow-up spectroscopic surveys of the fields of DLA absorbers revealed 
host galaxies with a wide range of morphological types between spirals, 
irregulars, dwarfs, and low surface brightness galaxies 
\citep[e.g.][]{lebrun97,rao03,chen05}.

Studies of elemental abundances in high-redshift DLAs have, on the other hand, 
achieved large statistical samples over the last few years. This progress is 
the result of 8--10\,m class telescopes and the high resolution spectroscopy. 
The HIRES spectrograph on the Keck~I telescope at Mauna Kea, Hawaii, was the 
first in operation and allowed the routine detection of $\approx 8$ elements in 
DLAs: Al, Si, S, Cr, Mn, Fe, Ni, and Zn \citep[e.g.][]{lu96,prochaska99}. With 
the arrival of the UVES spectrograph on the Very Large Telescope (VLT) ESO 
telescope at Cerro Paranal, Chile, with a better efficiency in the blue 
($\lambda < 4000$ \AA) and in the red ($\lambda > 7000$ \AA), $\approx 7$ 
additional element abundances can now be accurately measured in DLAs: C, N, O, 
Mg, P, Ar, and Ti, plus the deuterium and the molecular hydrogen
\citep[e.g.][]{levshakov02,petitjean00}. Moreover, for some specific DLAs, the 
so-called metal-strong DLAs, the access to $\approx 10$ new elements, several 
beyond the Fe-peak, is becoming very promising 
\citep{prochaska03,herbert-fort06}.

The main motivation common to all these observational efforts is to understand 
the physical processes at play in the formation and evolution of galaxies. To
achieve this goal, the chemical information can be used as one of the possible 
means at our disposal to link the properties of DLA high-redshift galaxies with
both those of present-day galaxies and other galaxies detected at high 
redshifts. Indeed, with the measure of so many different elemental abundances, 
DLAs are real physical laboratories. In 
\citet[][ hereafter Paper~I]{dessauges04}, we showed for three DLAs with 
comprehensive sets of elemental abundances (22 elements detected) that we can 
derive their star formation histories (SFHs) and ages from a detailed 
comparison of their abundance patterns with chemical evolution models. Previous
studies also investigated the nature of DLA galaxies by means of chemical 
evolution models \citep{jimenez99,hou01,cora03,calura03}. Our approach differs 
in the fact that we treat the DLAs studied as individual systems, and not as an 
ensemble following a common evolutionary sequence.

We propose to constrain the nature of six new high redshift DLA galaxies via 
their chemical abundance analysis following the prescriptions defined in 
Paper~I in order to derive a global picture of high-redshift galaxies. Together 
with the three DLAs previously studied, these nine DLA systems at $z_{\rm DLA} 
= 1.7-2.5$ were selected only on the basis of: (i)~a bright background quasar, 
(ii)~the existence of HIRES/Keck spectra, and (iii)~minimal Ly$\alpha$ forest 
contamination which maximizes access to chemical elements. The latter criterion 
was achieved by demanding $(z_{\rm QSO} - z_{\rm DLA}) < 0.6$ while requiring 
that the velocity separation $\Delta v({\rm QSO-DLA}) > 3000$ km~s$^{-1}$ such 
that the two are not physically associated. They form a sample which covers the 
full range of \ion{H}{i} column densities, metallicities, and dust depletion 
levels as observed in DLAs. The DLAs studied in this paper thus are well 
representative of the average high-redshift DLA galaxy population.
%without explicit bias to the metallicity, dust depletion level, or star 
%formation history; the sample only appears to be slightly biased toward high 
%\ion{H}{i} column densities. 

The layout of the paper is as follows. In Sect.~\ref{DLA-sample} we describe 
our DLA sample and the determination of their intrinsic abundances free from
dust depletion and ionization effects. In Sect.~\ref{models} we present the
chemical evolution models we will use to constrain the DLA star formation
histories and ages. In Sect.~\ref{SF-histories} we derive for the six DLA
galaxies in our sample the star formation histories and ages. Finally, in
Sect.~\ref{global-picture} we compare and try to connect the determined DLA 
properties, i.e. ages, star formation rates, and star formation histories, with 
the properties of emission-selected galaxies at the same redshifts. To 
conclude, Sect.~\ref{conclusions} provides a summary of the main constraints 
obtained on the DLA galaxies.  The adopted cosmology throughout the paper is 
$H_0 = 65$ km~s$^{-1}$~Mpc$^{-1}$ ($h = 0.65$), $\Omega_{\rm M} = 0.3$, and 
$\Omega_{\Lambda} = 0.7$. 

%
%________________________________________________________________

\section{Our DLA sample and their abundances}\label{DLA-sample}

The sample of DLA galaxies that we will study in this paper is unique. Indeed, 
by combining our high quality UVES/VLT spectra with existing HIRES/Keck 
spectra, we obtained comprehensive sets of elemental abundances for eleven 
DLAs. We detected 54 metal-line transitions which allowed us to measure the 
column densities of 30 ions from 22 elements: B, C, N, O, Mg, Al, Si, P, S, Cl, 
Ar, Ti, Cr, Mn, Fe, Co, Ni, Cu, Zn, Ge, As, and Kr. This contrasts with the 
majority of DLAs for which only a handful of elements (Si, Fe, Cr, Zn, Ni) is 
usually detected \citep[e.g.][]{lu96,prochaska99,prochaska01}. In these other 
cases, one struggles to interpret the DLA abundance patterns, especially to 
distinguish between differential depletion and intrinsic nucleosynthetic 
patterns \citep[e.g.][]{prochaska02b}.

%
%________________________________________________________________

\begin{sidewaystable*}
\caption{Summary of absolute and relative abundances} 
\label{abundances}
\begin{center}
{\scriptsize
\begin{tabular}{l l | l l | l l | l l | l l | l l | l l}
\hline\hline
\vspace{-0.2cm}\\
\multicolumn{2}{c}{Quasar}               & \multicolumn{2}{c}{Q\,B0841+129}              & \multicolumn{2}{c}{Q\,B0841+129}              & \multicolumn{2}{c}{PKS\,1157+014}             & \multicolumn{2}{c}{Q\,B1210+175}              & \multicolumn{2}{c}{Q\,B2230+02}               & \multicolumn{2}{c}{Q\,B2348--1444} 
\smallskip
\\ 
\hline
\vspace{-0.2cm}\\
\multicolumn{2}{c}{$z_{\rm DLA}$}        & \multicolumn{2}{c}{2.375}                     & \multicolumn{2}{c}{2.476\,$^{\dag}$}          & \multicolumn{2}{c}{1.944}                     & \multicolumn{2}{c}{1.892}                     & \multicolumn{2}{c}{1.864}                     & \multicolumn{2}{c}{2.279} \\
\multicolumn{2}{c}{$\log N$(\ion{H}{i})} & \multicolumn{2}{c}{20.99{\scriptsize (0.08)}} & \multicolumn{2}{c}{20.78{\scriptsize (0.08)}} & \multicolumn{2}{c}{21.60{\scriptsize (0.10)}} & \multicolumn{2}{c}{20.63{\scriptsize (0.08)}} & \multicolumn{2}{c}{20.83{\scriptsize (0.05)}} & \multicolumn{2}{c}{20.59{\scriptsize (0.08)}} 
\smallskip
\\
\hline
\vspace{-0.2cm}\\
\multicolumn{14}{l}{Ionic column densities and absolute abundances}
\smallskip
\\
\hline
\\[-0.2cm]	
$\log N$(\ion{B}{ii})  & [B/H]$_{\rm obs}$       & ...                       & ...                         & ...                       & ...                         & $< 11.93$		 & $< -0.46$		       & ...			   & ...			 & ...  		     & ...			   & ...		       & ...			     \\
$\log N$(\ion{N}{i})   & [N/H]$_{\rm obs}$       & 14.60{\scriptsize (0.01)} & $-2.29${\scriptsize (0.10)} & 13.94{\scriptsize (0.08)} & $-2.76${\scriptsize (0.13)} & $> 15.29$		 & $> -2.22$		       & 14.71{\scriptsize (0.09)} & $-1.82${\scriptsize (0.13)} & 15.02{\scriptsize (0.06)} & $-1.68${\scriptsize (0.11)} & 13.35{\scriptsize (0.05)} & $-3.16${\scriptsize (0.11)} \\
$\log N$(\ion{O}{i})   & [O/H]$_{\rm obs}$       & $> 16.05$                 & $> -1.77$                   & 16.15{\scriptsize (0.10)} & $-1.46${\scriptsize (0.14)} & ...			 & ...  		       & ...			   & ...			 & ...  		     & ...			   & $> 15.04$                 & $> -2.38$		     \\
$\log N$(\ion{Mg}{ii}) & [Mg/H]$_{\rm obs}$      & 15.13{\scriptsize (0.10)} & $-1.37${\scriptsize (0.13)} & 14.99{\scriptsize (0.15)} & $-1.35${\scriptsize (0.17)} & 15.99{\scriptsize (0.04)} & $-1.18${\scriptsize (0.11)} & $< 15.51$                 & $< -0.53$  		 & $< 15.61$		     & $< -0.80$		   & $< 14.80$  	       & $< -1.37$		     \\
                       & [Mg/H]$_{\rm cor}$(E00) &                           & $-1.37${\scriptsize (0.13)} &                           & ...                         &  			 & $-1.17${\scriptsize (0.12)} &			   & $< -0.53$  		 &			     & $< -0.77$		   &			       & $< -1.37$		     \\
		       & [Mg/H]$_{\rm cor}$(E11) &                           & $-1.37${\scriptsize (0.13)} &                           & ...                         &  			 & $-1.17${\scriptsize (0.12)} &			   & $< -0.53$  		 &			     & $< -0.77$		   &			       & $< -1.37$		     \\
$\log N$(\ion{Al}{ii}) & [Al/H]$_{\rm obs}$      & $> 13.72$                 & $> -1.76$                   & $> 13.37$                 & $> -1.90$                   & $> 15.01$                 & $> -1.08$		       & $> 14.94$                 & $> -0.18$			 & ...  		     & ...			   & $< 12.71$                 & $< -2.37$		     \\
$\log N$(\ion{Si}{ii}) & [Si/H]$_{\rm obs}$      & 15.21{\scriptsize (0.04)} & $-1.34${\scriptsize (0.09)} & 14.99{\scriptsize (0.03)} & $-1.35${\scriptsize (0.09)} & 15.97{\scriptsize (0.01)} & $-1.19${\scriptsize (0.10)} & 15.33{\scriptsize (0.03)} & $-0.86${\scriptsize (0.09)} & 15.70{\scriptsize (0.03)} & $-0.69${\scriptsize (0.06)} & 14.18{\scriptsize (0.06)} & $-1.97${\scriptsize (0.10)} \\
                       & [Si/H]$_{\rm cor}$(E00) &                           & $-1.34${\scriptsize (0.09)} &                           & ...                         &  			 & $-1.19${\scriptsize (0.10)} &			   & $-0.86${\scriptsize (0.09)} &			     & $-0.68${\scriptsize (0.07)} &			       & $-1.96${\scriptsize (0.10)} \\
		       & [Si/H]$_{\rm cor}$(E11) &                           & $-1.34${\scriptsize (0.09)} &                           & ...                         &  			 & $-1.18${\scriptsize (0.11)} &			   & $-0.86${\scriptsize (0.09)} &			     & $-0.68${\scriptsize (0.07)} &			       & $-1.96${\scriptsize (0.10)} \\
$\log N$(\ion{P}{ii})  & [P/H]$_{\rm obs}$       & 12.82{\scriptsize (0.06)} & $-1.73${\scriptsize (0.12)} & 12.56{\scriptsize (0.09)} & $-1.78${\scriptsize (0.13)} & 13.86{\scriptsize (0.09)} & $-1.30${\scriptsize (0.15)} & $< 13.95$                 & $< -0.24$			 & 13.69{\scriptsize (0.08)} & $-0.70${\scriptsize (0.11)} & ...		       & ...			     \\
$\log N$(\ion{S}{ii})  & [S/H]$_{\rm obs}$       & 14.69{\scriptsize (0.05)} & $-1.48${\scriptsize (0.11)} & 14.48{\scriptsize (0.10)} & $-1.50${\scriptsize (0.15)} & ...			 & ...  		       & 14.96{\scriptsize (0.02)} & $-0.85${\scriptsize (0.10)} & 15.29{\scriptsize (0.06)} & $-0.66${\scriptsize (0.11)} & 13.75{\scriptsize (0.06)} & $-2.03${\scriptsize (0.12)} \\
$\log N$(\ion{Cl}{i})  & [Cl/H]$_{\rm obs}$      & ...                       & ...                         & ...                       & ...                         & $< 12.81$		 & $< -2.07$    	       & ...			   & ...			 & ...  		     & ...			   & ...		       & ...			     \\
$\log N$(\ion{Ar}{i})  & [Ar/H]$_{\rm obs}$      & 13.53{\scriptsize (0.08)} & $-1.86${\scriptsize (0.13)} & 13.11{\scriptsize (0.08)} & $-2.07${\scriptsize (0.13)} & ...			 & ...  		       & ...			   & ...			 & ...  		     & ...			   & ...		       & ...			     \\
$\log N$(\ion{Ti}{ii}) & [Ti/H]$_{\rm obs}$      & ...                       & ...                         & ...                       & ...                         & 12.82{\scriptsize (0.13)} & $-1.71${\scriptsize (0.16)} & 12.34{\scriptsize (0.08)} & $-1.06${\scriptsize (0.13)} & 12.68{\scriptsize (0.09)} & $-1.09${\scriptsize (0.10)} & ...		       & ...			     \\
$\log N$(\ion{Cr}{ii}) & [Cr/H]$_{\rm obs}$      & 13.07{\scriptsize (0.02)} & $-1.59${\scriptsize (0.08)} & 12.89{\scriptsize (0.06)} & $-1.58${\scriptsize (0.11)} & 13.76{\scriptsize (0.02)} & $-1.52${\scriptsize (0.10)} & 13.28{\scriptsize (0.03)} & $-1.02${\scriptsize (0.09)} & 13.44{\scriptsize (0.04)} & $-0.97${\scriptsize (0.08)} & 12.30{\scriptsize (0.09)} & $-1.97${\scriptsize (0.12)} \\
%                      & [Cr/H]$_{\rm cor}$(E00) &                           & $-1.51${\scriptsize (0.09)} &                           & ...                         &  			 & $-1.34${\scriptsize (0.16)} &			   & $-0.94${\scriptsize (0.14)} &			     & $-0.74${\scriptsize (0.15)} &			       & $-1.93${\scriptsize (0.16)} \\
%		       & [Cr/H]$_{\rm cor}$(E11) &                           & $-1.51${\scriptsize (0.09)} &                           & ...                         &  			 & $-1.33${\scriptsize (0.17)} &			   & $-0.94${\scriptsize (0.14)} &			     & $-0.73${\scriptsize (0.15)} &			       & $-1.90${\scriptsize (0.19)} \\
$\log N$(\ion{Mn}{ii}) & [Mn/H]$_{\rm obs}$      & 12.50{\scriptsize (0.02)} & $-2.02${\scriptsize (0.08)} & 12.33{\scriptsize (0.15)} & $-1.98${\scriptsize (0.17)} & 13.25{\scriptsize (0.02)} & $-1.87${\scriptsize (0.10)} & 12.72{\scriptsize (0.01)} & $-1.42${\scriptsize (0.09)} & 13.03{\scriptsize (0.03)} & $-1.33${\scriptsize (0.06)} & 11.66{\scriptsize (0.15)} & $-2.46${\scriptsize (0.17)} \\
$\log N$(\ion{Fe}{ii}) & [Fe/H]$_{\rm obs}$      & 14.76{\scriptsize (0.01)} & $-1.73${\scriptsize (0.08)} & 14.50{\scriptsize (0.03)} & $-1.78${\scriptsize (0.09)} & 15.46{\scriptsize (0.02)} & $-1.64${\scriptsize (0.10)} & 15.01{\scriptsize (0.03)} & $-1.12${\scriptsize (0.09)} & 15.24{\scriptsize (0.03)} & $-1.09${\scriptsize (0.06)} & 13.84{\scriptsize (0.05)} & $-2.25${\scriptsize (0.09)} \\
                       & [Fe/H]$_{\rm cor}$(E00) &                           & $-1.59${\scriptsize (0.10)} &                           & ...                         &  			 & $-1.36${\scriptsize (0.18)} &			   & $-0.98${\scriptsize (0.16)} &			     & $-0.76${\scriptsize (0.14)} &			       & $-2.13${\scriptsize (0.19)} \\
		       & [Fe/H]$_{\rm cor}$(E11) &                           & $-1.59${\scriptsize (0.10)} &                           & ...                         &  			 & $-1.36${\scriptsize (0.18)} &			   & $-0.98${\scriptsize (0.16)} &			     & $-0.75${\scriptsize (0.14)} &			       & $-2.13${\scriptsize (0.19)} \\
$\log N$(\ion{Co}{ii}) & [Co/H]$_{\rm obs}$      & ...                       & ...                         & ...                       & ...                         & 13.10{\scriptsize (0.13)} & $-1.40${\scriptsize (0.16)} & ...			   & ...			 & ...  		     & ...			   & ...		       & ...			     \\
$\log N$(\ion{Ni}{ii}) & [Ni/H]$_{\rm obs}$      & 13.53{\scriptsize (0.05)} & $-1.69${\scriptsize (0.09)} & 13.19{\scriptsize (0.09)} & $-1.84${\scriptsize (0.13)} & 14.21{\scriptsize (0.02)} & $-1.66${\scriptsize (0.10)} & 13.67{\scriptsize (0.06)} & $-1.19${\scriptsize (0.11)} & 14.09{\scriptsize (0.04)} & $-0.92${\scriptsize (0.08)} & $< 12.30$  	       & $< -2.30$		     \\
                       & [Ni/H]$_{\rm cor}$(E00) &                           & $-1.56${\scriptsize (0.11)} &                           & ...                         &  			 & $-1.37${\scriptsize (0.18)} &			   & $-1.03${\scriptsize (0.19)} &			     & $-0.67${\scriptsize (0.13)} &			       & $< -2.17$		     \\
		       & [Ni/H]$_{\rm cor}$(E11) &                           & $-1.55${\scriptsize (0.11)} &                           & ...                         &  			 & $-1.38${\scriptsize (0.18)} &			   & $-1.05${\scriptsize (0.18)} &			     & $-0.59${\scriptsize (0.16)} &			       & $< -2.18$		     \\
$\log N$(\ion{Cu}{ii}) & [Cu/H]$_{\rm obs}$      & ...                       & ...                         & ...                       & ...                         & $< 12.31$		 & $< -1.56$		       & ...			   & ...			 & ...  		     & ...			   & ...		       & ...			     \\
$\log N$(\ion{Zn}{ii}) & [Zn/H]$_{\rm obs}$      & 12.10{\scriptsize (0.02)} & $-1.49${\scriptsize (0.09)} & 11.69{\scriptsize (0.10)} & $-1.74${\scriptsize (0.14)} & 12.99{\scriptsize (0.05)} & $-1.27${\scriptsize (0.12)} & 12.40{\scriptsize (0.05)} & $-0.88${\scriptsize (0.10)} & 12.72{\scriptsize (0.05)} & $-0.67${\scriptsize (0.09)} & $< 11.28$  	       & $<-1.97$		     \\
$\log N$(\ion{Ge}{ii}) & [Ge/H]$_{\rm obs}$      & ...                       & ...                         & ...                       & ...                         & 11.99{\scriptsize (0.14)} & $-1.24${\scriptsize (0.18)} & ...			   & ...			 & ...  		     & ...			   & ...		       & ...			     \\
$\log N$(\ion{As}{ii}) & [As/H]$_{\rm obs}$      & ...                       & ...                         & ...                       & ...                         & $< 11.78$		 & $< -0.19$		       & $< 11.97$		   & $< +0.97$			 & ...  		     & ...			   & ...		       & ...			     \\
$\log N$(\ion{Kr}{i})  & [Kr/H]$_{\rm obs}$      & ...                       & ...                         & ...                       & ...                         & $< 13.01$		 & $< +0.10$		       & $< 12.48$		   & $< +0.54$			 & ...  		     & ...			   & ...		       & ...			     \\
$\log N$(\ion{C}{ii}$^*$) & --                   & 12.96{\scriptsize (0.08)} & --                          & $< 13.08$                 & --                          & 15.13{\scriptsize (0.09)} & --   		       & ...			   & -- 			 & 13.95{\scriptsize (0.02)} & -- 			   & $< 12.25$                 & -- 			     
\\[0.1cm]
\hline
\vspace{-0.2cm}\\
\multicolumn{14}{l}{Relative abundances} 
\smallskip
\\
\hline
\\[-0.2cm]
 & [Zn/Fe]$_{\rm obs}$       & & $+0.24${\scriptsize (0.02)} & & $+0.04${\scriptsize (0.11)} & & $+0.37${\scriptsize (0.07)} & & $+0.24${\scriptsize (0.07)} & & $+0.42${\scriptsize (0.07)} & & $< +0.28$                   \\
 & [Ni/Fe]$_{\rm obs}$       & & $+0.04${\scriptsize (0.05)} & & $-0.06${\scriptsize (0.09)} & & $-0.02${\scriptsize (0.03)} & & $-0.07${\scriptsize (0.07)} & & $+0.17${\scriptsize (0.05)} & & $< -0.05$                   \\
 & [Ni/Fe]$_{\rm cor}$ (E00) & & $+0.03${\scriptsize (0.04)} & & ...                         & & $-0.01${\scriptsize (0.02)} & & $-0.05${\scriptsize (0.06)} & & $+0.09${\scriptsize (0.04)} & & $< -0.04$                   \\
 & [Ni/Fe]$_{\rm cor}$ (E11) & & $+0.04${\scriptsize (0.05)} & & ...                         & & $-0.02${\scriptsize (0.03)} & & $-0.07${\scriptsize (0.07)} & & $+0.16${\scriptsize (0.05)} & & $< -0.05$                   \\
 & [Si/Fe]$_{\rm obs}$       & & $+0.39${\scriptsize (0.04)} & & $+0.43${\scriptsize (0.04)} & & $+0.45${\scriptsize (0.02)} & & $+0.26${\scriptsize (0.03)} & & $+0.40${\scriptsize (0.04)} & & $+0.28${\scriptsize (0.08)} \\
 & [Si/Fe]$_{\rm cor}$ (E00) & & $+0.25${\scriptsize (0.06)} & & ...                         & & $+0.18${\scriptsize (0.09)} & & $+0.12${\scriptsize (0.10)} & & $+0.08${\scriptsize (0.11)} & & $+0.16${\scriptsize (0.18)} \\
 & [Si/Fe]$_{\rm cor}$ (E11) & & $+0.25${\scriptsize (0.06)} & & ...                         & & $+0.18${\scriptsize (0.09)} & & $+0.12${\scriptsize (0.10)} & & $+0.08${\scriptsize (0.11)} & & $+0.16${\scriptsize (0.18)} \\
 & [Mg/Fe]$_{\rm obs}$       & & $+0.36${\scriptsize (0.10)} & & $+0.43${\scriptsize (0.15)} & & $+0.46${\scriptsize (0.05)} & & $< +0.59$                   & & $< +0.29$                   & & $< +0.88$                   \\
 & [Mg/Fe]$_{\rm cor}$ (E00) & & $+0.22${\scriptsize (0.12)} & & ...                         & & $+0.19${\scriptsize (0.12)} & & $< +0.45$                   & & $< -0.02$                   & & $< +0.70$                   \\
 & [Mg/Fe]$_{\rm cor}$ (E11) & & $+0.22${\scriptsize (0.12)} & & ...                         & & $+0.19${\scriptsize (0.12)} & & $< +0.45$                   & & $< -0.02$                   & & $< +0.70$                   \\
 & [S/Fe]$_{\rm obs}$        & & $+0.25${\scriptsize (0.08)} & & $+0.28${\scriptsize (0.12)} & & ...                         & & $+0.27${\scriptsize (0.07)} & & $+0.43${\scriptsize (0.09)} & & $+0.22${\scriptsize (0.10)} \\
 & [S/Fe]$_{\rm cor}$ (E00)  & & $+0.11${\scriptsize (0.13)} & & ...                         & & ...                         & & $+0.13${\scriptsize (0.10)} & & $+0.10${\scriptsize (0.18)} & & $+0.10${\scriptsize (0.18)} \\
 & [S/Fe]$_{\rm cor}$ (E11)  & & $+0.11${\scriptsize (0.13)} & & ...                         & & ...                         & & $+0.13${\scriptsize (0.10)} & & $+0.09${\scriptsize (0.18)} & & $+0.10${\scriptsize (0.18)} \\
 & [S/Zn]$_{\rm obs}$        & & $+0.02${\scriptsize (0.07)} & & $+0.23${\scriptsize (0.15)} & & ...                         & & $+0.03${\scriptsize (0.08)} & & $+0.00${\scriptsize (0.10)} & & $> -0.06$                   \\
 & [N/Si]$_{\rm obs}$        & & $-0.95${\scriptsize (0.07)} & & $-1.41${\scriptsize (0.10)} & & $> -1.03$                   & & $-0.96${\scriptsize (0.11)} & & $-0.98${\scriptsize (0.09)} & & $-1.17${\scriptsize (0.10)} \\
 & [N/Si]$_{\rm cor}$ (E00)  & & $-0.95${\scriptsize (0.07)} & & ...                         & & $> -1.03$                   & & $-0.96${\scriptsize (0.16)} & & $-1.00${\scriptsize (0.13)} & & $-1.18${\scriptsize (0.10)} \\
 & [N/Si]$_{\rm cor}$ (E11)  & & $-0.95${\scriptsize (0.07)} & & ...                         & & $> -1.04$                   & & $-0.96${\scriptsize (0.16)} & & $-1.00${\scriptsize (0.13)} & & $-1.18${\scriptsize (0.10)} \\
 & [N/S]$_{\rm obs}$         & & $-0.81${\scriptsize (0.07)} & & $-1.26${\scriptsize (0.14)} & & ...                         & & $-0.97${\scriptsize (0.11)} & & $-1.01${\scriptsize (0.10)} & & $-1.12${\scriptsize (0.10)} 
\\[0.1cm]
\hline
\end{tabular}}
\begin{minipage}{220mm}
\smallskip
{\scriptsize
{\it Notes.} -- The DLA at $z_{\rm DLA} = 2.067$ toward Q\,B0450$-$1310 is affected by high ionization 
effects (see Sect.~\ref{ionization}), and therefore we do not consider it in this paper. \\
\phantom{{\it Notes.} } -- The values in parenthesis correspond to the 1~$\sigma$ errors. \\
\phantom{{\it Notes.} } -- We provide dust corrections only for the absolute and relative abundances 
that we will further use in our study. The dust corrections are computed using the \citet{vladilo02a} 
prescriptions. The models E00 and E11 \\
\phantom{{\it Notes.} } assume that the intrinsic [Zn/Fe] ratio is equal to $+0.10$~dex. \\
$^{\dag}$ The DLA at $z_{\rm DLA} = 2.476$ toward Q\,B0841+129 does not require any dust correction, 
because the [Zn/Fe] ratio is almost solar. We obtained for this DLA the rare O abundance measurement 
which leads to the following relative abundances: [O/Fe] $= +0.32\pm 0.12$, [O/Zn] $= +0.26\pm 0.15$, 
and [N/O] $= -1.30\pm 0.14$. }
\end{minipage}
\end{center}
\end{sidewaystable*}		
%
%________________________________________________________________

The abundance analysis of our sample of DLAs is presented in Paper~I for four 
DLAs toward Q\,B0100+13, Q\,B1331+17, Q\,B2231$-$002, and Q\,B2343+125, and in 
\citet[][ hereafter Paper~II]{dessauges06} for seven DLAs toward 
Q\,B0450$-$1310, Q\,B0841+129, PKS\,1157+014, Q\,B1210+175, Q\,B2230+02, and 
Q\,B2348$-$1444\footnote{The naming of quasars have slightly been modified
relative to Papers~I and II in order to be in accordance with the IAU rules.}. 
A complete interpretation of the abundances with the determination of the star 
formation histories can also be found in Paper~I for the first set of 
DLAs\footnote{The DLA toward Q\,B2343+125 was excluded from the star formation 
history study, because of its large ionization fraction.}. In this paper we 
will focus on the second set of DLAs. In Table~\ref{abundances} we summarize 
their ionic column densities and absolute abundances. The most important 
relative abundances are also reported. They have been computed by considering 
only the column densities of Voigt profile components detected in both elements 
of the abundance ratio.

The interpretation of the observed elemental abundance patterns in DLAs is not
straightforward. The principal difficulty is to disentangle the nucleosynthesis
contributions from dust depletion and ionization effects. Indeed, since we are
measuring gas-phase elemental abundances, in presence of dust the observed
abundances may not represent the intrinsic chemical composition of the galaxy 
if part of the elements is removed from the gas to the solid phase, as it
happens in the interstellar medium (ISM) of our Galaxy \citep{savage96}. 
Similarly, a fraction of the gas may be ionized by external or internal 
ionizing sources. Therefore, a comprehensive set of ionic column densities and
elemental abundances is of great advantage in order to evaluate the dust 
depletion and ionization effects, and hence derive the intrinsic abundances. 

%
%________________________________________________________________

\subsection{Dust depletion corrections}\label{dust-depletion}

Several pieces of evidence show that some dust is present in DLAs
\citep{pettini94,hou01,prochaska02b}. Three main approaches help to circumvent 
the problem of dust depletion in the studies of DLA abundance patterns. The 
first one, although very limiting, consists in considering only the DLAs with 
low dust depletion levels. [Zn/Fe], the ratio of two iron-peak elements with 
very different degrees of incorporation into dust grains, is often used as a 
dust depletion indicator, with Zn being a volatile element and Fe a refractory 
element. Hence, DLAs with [Zn/Fe] $\lesssim 0.2$ are considered as systems with 
negligible dust depletion levels \citep{pettini00a}. This is the case in our 
sample for the DLA at $z_{\rm DLA} = 2.476$ toward Q\,B0841+129 with [Zn/Fe] 
$= +0.04\pm 0.11$. In three other DLAs toward Q\,B0841+129, Q\,B1210+175, and 
Q\,B2348$-$1444, the dust depletion levels are also relatively low: [Zn/Fe] 
$< +0.28$. The second approach consists of quantifying the dust depletion 
effects and applying dust corrections to the observed abundances. Several 
authors have considered different methods to compute the dust depletion 
corrections 
\citep[e.g.][]{vladilo98,savaglio00,vladilo02a,vladilo02b,prochaska02b}. 
Finally, the third approach consists of focusing on non-refractory and/or 
mildly refractory elements, such as N, O, S, Ar, and Zn. In this way one 
assesses the intrinsic abundances in DLAs.

In our study we considered all the three approaches: the first for the DLA at 
$z_{\rm DLA} = 2.476$ toward Q\,B0841+129, and the second and the third for the 
other DLAs. We performed the dust depletion corrections using the method 
developed by \citet{vladilo02a,vladilo02b}, as in Paper~I. This method groups 
together several dust correction models based on different assumptions. We
consider here the models E00 and E11 computed by assuming that the intrinsic 
[Zn/Fe] ratio is equal to +0.10~dex (see Vladilo's papers for more details). In 
Table~\ref{abundances} we report the dust-corrected values, in addition to the 
observed ones, for the absolute and relative abundances which we will further 
use in the paper. 

%
%________________________________________________________________

\subsection{Ionization corrections}\label{ionization}

Several authors investigated photoionization in DLAs
\citep[e.g.][]{viegas95,howk99,izotov01,vladilo01} and generally concluded that 
the ionization corrections to observed abundances should be small. We used the 
photoionization diagnostics of \citet{prochaska02a} to check the ionization 
state of the DLAs in our sample. These diagnostics provide a qualitative 
analysis of the ionization level in a DLA. They consist of prescriptions based 
on the column density ratios of adjacent ions of the same element, such as 
$N$(Fe$^{++}$)/$N$(Fe$^+$), and on Ar. Ar is very sensitive to ionization, 
because its photoionization cross-section is one order of magnitude larger than 
for \ion{H}{i} at energies $h\nu > 2$~Ryd \citep{sofia98}. 

These photoionization diagnostics were discussed in Papers~I and II for the 
DLAs in our sample. They show that the ionization fraction is lower than 10\,\% 
for nine out of eleven DLAs and that the ionization corrections are negligible 
for these nine DLAs. The two DLAs with the high ionization fractions, the DLA
toward Q\,B2343+125 and the DLA toward Q\,B0450$-$1310, are not considered in 
the star formation study and therefore the DLA toward Q\,B0450$-$1310 is not 
included in Table~\ref{abundances}. They will be discussed in a future paper
(Deassuages-Zavadsky et~al.\ in prep).

%
%______________________________________________________________

\begin{figure*}[t]
\centering
\includegraphics[width=8.5cm]{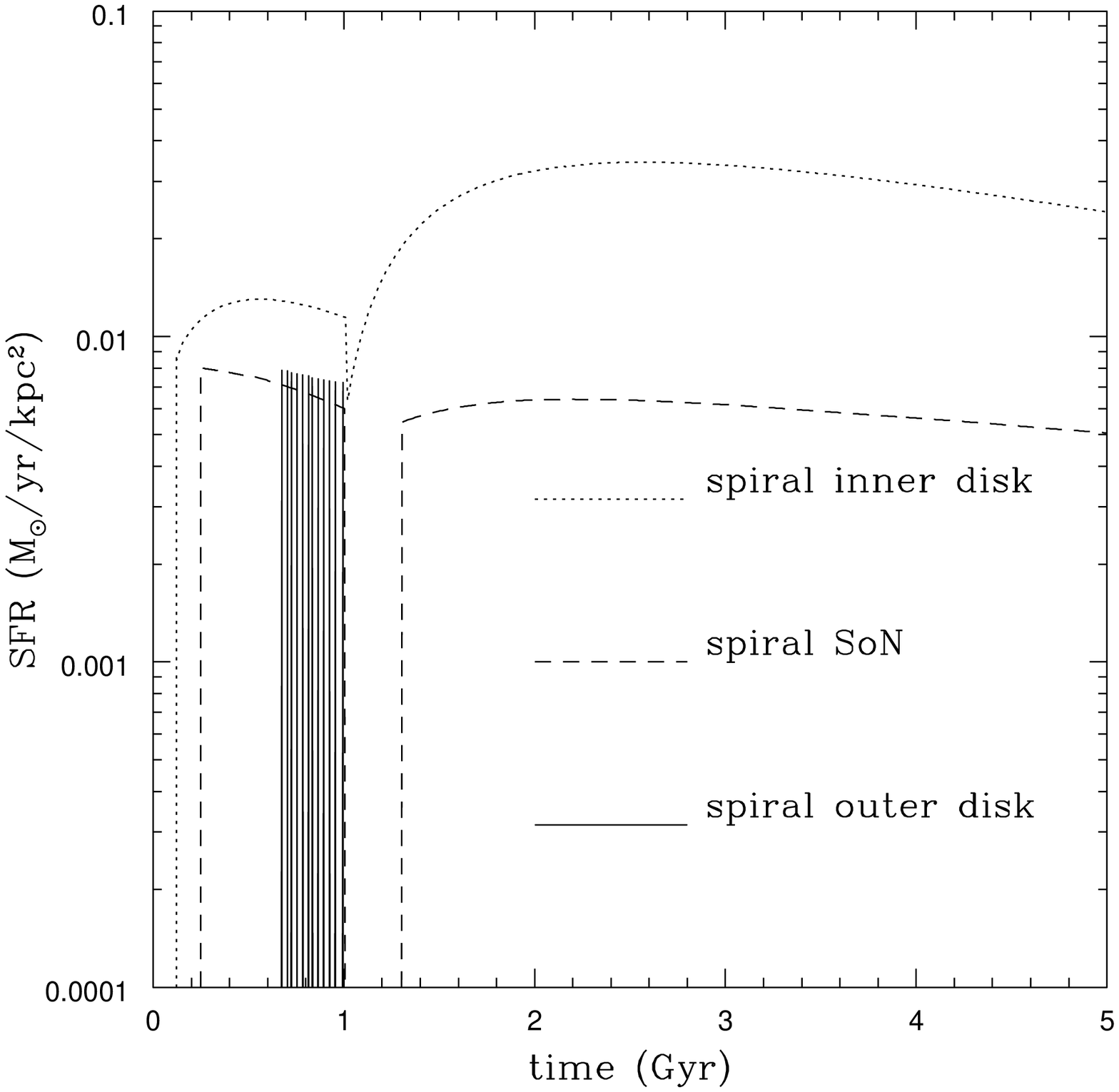}\hfill
\includegraphics[width=8.5cm]{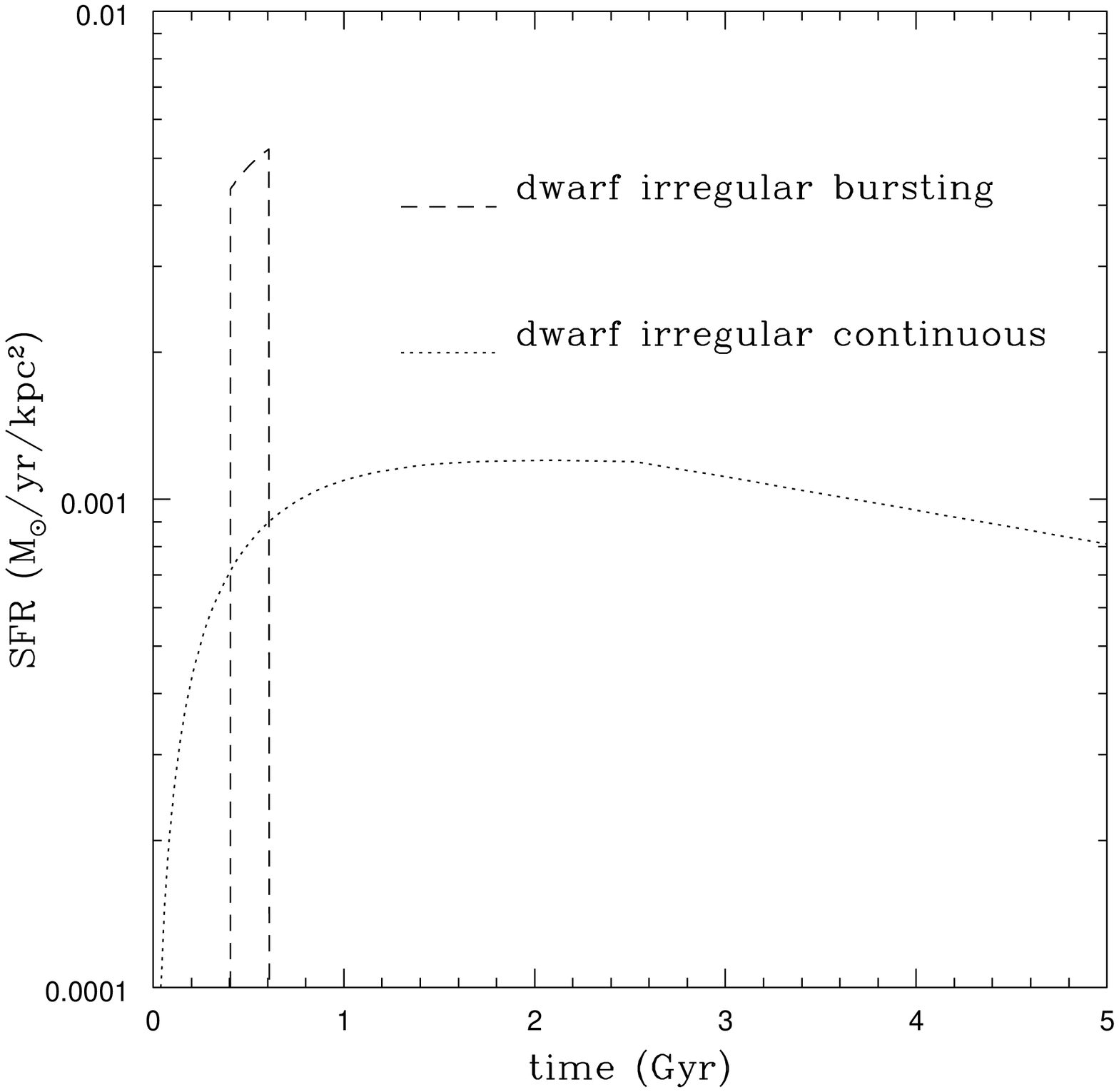}
\caption{{\it Left panel.} Star formation rates per unit area as a function of 
time of the spiral models at different galactocentric radii. In the inner 
regions of the spiral disk ($R < 6$ kpc) the star formation proceeds 
continuously (dotted line), in the solar neighborhood (SoN, $R\sim 8$ kpc) the 
star formation also proceeds continuously but at a weaker efficiency (dashed 
line), and finally in the outer regions of the spiral disk ($R > 10$ kpc) the 
star formation proceeds in a bursting way (solid line) due to the gas surface 
density threshold adopted for the star formation.
{\it Right panel.} Star formation rates per unit area as a function of time of 
the dwarf irregular models. Models with two different star formation regimes 
are illustrated: the dwarf irregular model with a continuous star formation 
(dotted line) and the dwarf irregular model with a bursting star formation 
(dashed line). The dwarf irregular continuous models are characterized by about 
10 times weaker star formation rates than the spiral inner disk and solar 
neighborhood models, and the dwarf irregular bursting models are characterized 
by a single burst of star formation. Here we show a burst lasting 200 Myr and 
having an efficiency of 0.3 Gyr$^{-1}$, a model which turned out to well 
represent the star formation history of some DLAs studied in Paper~I.}
\label{SFH-spiral-dwarf}
\end{figure*}
%JXP -- I think it might make more sense to show this Figure as a plot vs. nu
%  not SFR/area.  This is especially true for the dwarf models.  Or, perhaps the
%  right-hand axis could be labeled with nu?
%
%______________________________________________________________

\section{Chemical evolution models}\label{models}

Theoretically, we can determine the star formation history and the age of a
galaxy from its abundance patterns \citep{matteucci01}. Indeed, in chemical
evolution models the absolute abundances depend on the model assumptions,
whereas the relative abundances depend on the nucleosynthesis, the stellar
lifetimes, and the initial mass function. Relative abundances can thus be used 
as cosmic clocks if they involve two elements formed on different timescales. 
Typical examples are the [$\alpha$/Fe] and [N/$\alpha$] ratios, since the 
$\alpha$-elements (O, Si, S, Mg) are produced on short timescales by Type~II 
supernovae (SNe) and the Fe-peak elements and nitrogen are produced on long 
timescales by Type~Ia SNe and intermediate-mass stars, respectively. 

In this work, we will make use of different chemical evolution models spanning
different star formation histories and star formation efficiencies. These 
models are the same as the models used by \citet{calura03} and in Paper~I. No 
instantaneous recycling approximation is adopted, the stellar lifetimes are 
properly taken into account. We describe the models in the next two sections.

%
%________________________________________________________________

\subsection{Spiral models}\label{spiral-model}

The model for a typical spiral galaxy is calibrated on the chemical features of
the Milky Way. The galaxy is assumed to form as a result of two main infall 
episodes \citep{chiappini97}. During the first episode the halo forms and the 
gas shed by the halo rapidly gathers in the center leading to the formation of 
the bulge. During the second episode, a slower infall of external gas gives 
rise to the disk with the gas accumulating faster in the inner than in the 
outer region, leading to timescales ranging from $\sim 2$ Gyr in the inner disk 
to $\sim 7$ Gyr in the solar region, and up to 13 Gyr in the outer disk
\citep[``inside-out'' scenario,][]{matteucci89}. The efficiency of the star
formation is characterized by the efficiency parameter, $\nu$, defined as the 
inverse of the typical timescale for star formation. It is set to $\nu = 1$ 
Gyr$^{-1}$ and becomes zero when the gas surface density drops below a certain 
critical threshold. A threshold density of $\sim 7$ M$_{\odot}$~pc$^{-2}$ is 
adopted in the disk \citep{chiappini97}. The resulting star formation rate then 
is proportional to $\nu$ and the gas surface density which depends on the 
galactocentric radius $R$ and time. The initial mass function (IMF) is taken 
from \citet{scalo86}. The galactic wind is assumed to be inefficient 
\citep{matteucci01}. 

Hence, due to the ``inside-out'' scenario and the threshold density adopted in
the spiral model, the star formation history is different at different 
galactocentric radii (see Fig.~\ref{SFH-spiral-dwarf}, left panel). Indeed, in 
the inner regions of the disk where the rate of accretion of matter onto the 
disk is fast, a high gas surface density well above the critical threshold is 
reached and maintained during a long period, and hence the star formation is 
almost continuous through the galaxy lifetime. On the other hand, in the outer 
regions of the disk (radius $R \gtrsim 10$ kpc) where the rate of accretion of 
matter onto the disk is slow, the SFH proceeds in a gasping way, due to the 
fact that in these regions the gas density is always close to the critical 
threshold. We will label ``spiral inner disk'' the models at $R < 6$ kpc, 
``spiral solar neighborhood'' the model at $R\sim 8$ kpc, and ``spiral outer 
disk'' the models at $R > 10$ kpc. These different star formation histories at 
different galactocentric radii are associated with different abundance patterns 
which can be compared with observations. The radius $R$ corresponds to the 
position at which the QSO line of sight crosses the DLA galaxy disk.

%
%________________________________________________________________

\subsection{Dwarf irregular models}\label{dwarf-model}

Dwarf irregular galaxies are assumed to form owing to a continuous infall of 
pristine gas until a mass of $\sim 10^9$ M$_{\odot}$ is accumulated 
\citep{bradamante98}. The evolution of these galaxies is characterized by a 
bursting star formation history. The parameters which need to be constrained 
are the number of bursts of star formation and for each burst the efficiency, 
$\nu$, (same definition as above) and the burst duration, $\Delta t$. The star 
formation rate is in this case directly proportional to $\nu$ and the gas 
density at a given time. The star formation in the dwarf irregular models can 
proceed either in short bursts of a duration from 10 to 200 Myr separated by 
long quiescent periods or at a low regime but continuously, namely in one or 
two long episodes of inefficient star formation lasting between 3 and 13 Gyr 
(see Fig.~\ref{SFH-spiral-dwarf}, right panel). They will be referred as the 
``dwarf irregular bursting'' models and the ``dwarf irregular continuous'' 
models, respectively. The dwarf irregular galaxies are particularly sensitive 
to outflows resulting from the energy injection from the supernova explosions 
\citep{recchi01}. 

The IMF is taken from \citet{salpeter55}. The reason for such a choice relies 
mainly on the chemical abundances and on the metal content observed in 
irregulars \citep[for a detailed explanation, see][]{calura06}. Indeed, a 
Salpeter IMF can account for the abundances observed in local dwarf galaxies 
\citep{recchi02}, whereas it leads to an overestimation of the abundances in 
spiral disks \citep[see][]{romano05} which are well accounted for by means of 
the \citet{scalo86} IMF. On the other hand, when assuming a Scalo IMF in 
dwarfs, we underestimate their present-day average metallicity 
\citep{calura04}. 

%
%________________________________________________________________

\vspace{0.6cm}

\noindent In summary, three different types of star formation histories are 
modeled by the grid of chemical evolution models which we use in our study of 
DLAs (see Fig.~\ref{SFH-spiral-dwarf} for illustration): 
\begin{itemize}
\item the single burst SFH: \\
by the {\it dwarf irregular bursting models}: dwarf irregular models with a 
single burst of star formation characterized by different efficiencies $\nu$ 
from 0.1 to 5 Gyr$^{-1}$ and different burst durations $\Delta t$ from 10 to 
200 Myr;
\item the episodic bursting SFH: \\
by the {\it spiral outer disk models}: spiral models at $R > 10$ kpc 
characterized by a bursting star formation separated by short quiescent 
periods; 
\item the continuous SFH: \\
by the {\it spiral inner disk models}: spiral models at $R < 6$ kpc 
characterized by a continuous star formation over 13 Gyr; \\ 
by the {\it spiral solar neighborhood models}: spiral models at $R\sim 8$ kpc 
characterized by a continuous star formation but having a weaker star formation
efficiency than the spiral inner disk models; \\
and by the {\it dwarf irregular continuous models}: dwarf irregular models with a 
continuous star formation over $\Delta t = 13$ Gyr and characterized by low 
efficiencies $\nu$ from 0.01 to 0.1 Gyr$^{-1}$, leading to star formation
efficiencies about 10 times weaker than in the spiral inner disk and solar 
neighborhood models.
\end{itemize}

%
%________________________________________________________________

\begin{table*}[t]
\begin{center}
\caption{Star formation histories, ages, and star formation rates per unit area 
of the DLA galaxies studied} 
\label{SFH}
\begin{tabular}{l c l l c c c}
\hline \hline
\\[-0.3cm]
QSO & $z_{\rm DLA}$ & Possible model & Star formation characteristics          & $z_f$ & Age   & SFR per unit area \\
    &               &                & $\nu$ [Gyr$^{-1}$]~/~$\Delta t$ [Gyr] &       & [Gyr] & [M$_{\odot}$~yr$^{-1}$~kpc$^{-2}$] 
\smallskip 
\\     
\hline  
\\[-0.3cm]
Q\,B0841+129    & 2.375 & spiral                     & outer disk                & $\sim 2.7$       & $0.2-0.4$       & $7.5\times 10^{-3}$ \\
                &       & {\bf dwarf irregular}      & {\bf bursting: 0.3~/~0.2} & ${\bf 2.6-3.0}$  & ${\bf 0.2-0.6}$ & ${\bf 4.0\times 10^{-3}}$ \\
	        &       & dwarf irregular            & continuous: 0.05          & $\sim 4.0$       & $1.0-1.5$       & $1.2\times 10^{-3}$ \\  
\\

Q\,B0841+129    & 2.476 & {\bf dwarf irregular}      & {\bf bursting: 2.0~/~0.02}& ${\bf \sim 2.5}$ & ${\bf 0.01-0.03}$ & ${\bf 3.0\times 10^{-2}}$ \\
\\

PKS\,1157+014   & 1.944 & {\bf spiral}               & {\bf outer disk}          & ${\bf 2.0-2.4}$  & ${\bf 0.1-0.6}$ & ${\bf 7.5\times 10^{-3}}$ \\
                &       & dwarf irregular	     & bursting: 0.7~/~0.1       & $2.4-2.9$        & $0.7-1.2$       & $1.6\times 10^{-2}$ \\
	        &       & dwarf irregular	     & continuous: 0.05	         & $4.0-5.5$        & $1.9-2.5$       & $7.5\times 10^{-4} - 1.0\times 10^{-3}$ \\
\\

Q\,B1210+175    & 1.892 & spiral                     & inner disk                & $1.9-2.2$        & $0.1-0.5$       & $8.0\times 10^{-3} - 1.3\times 10^{-2}$ \\
                &       & {\bf dwarf irregular}      & {\bf bursting: 1.2~/~0.2} & ${\bf 2.2-2.6}$  & ${\bf 0.5-1.0}$ & ${\bf 1.7\times 10^{-2}}$ \\
	        &       & dwarf irregular            & continuous: 0.06          & $> 5$            & $> 2.4$         & $8.0\times 10^{-4}$ \\
\\

Q\,B2230+02     & 1.864 & {\bf dwarf irregular}      & {\bf continuous: 0.1}     & ${\bf > 10}$     & ${\bf > 3.2}$   & ${\bf 6.5\times 10^{-4}}$ \\
\\

Q\,B2348$-$1444 & 2.279 & {\bf dwarf irregular}      & {\bf bursting: 0.1~/~0.15}& ${\bf 2.4-2.5}$  & ${\bf 0.1-0.3}$ & ${\bf 2.2\times 10^{-3}}$ \\
             &       & dwarf irregular       & continuous: 0.1           & $2.6-2.7$        & $0.3-0.5$       & $8.7\times 10^{-4}$ \\[0.05cm]
\hline
\end{tabular}
\begin{minipage}{165mm}
\smallskip
{\it Note.} The model in boldface corresponds to the {\it favored} model. We assume that 
the favored model is the model which reproduces \\
\phantom{{\it Note.}} all the data points within 1~$\sigma$ error (minimum $\chi^2$). 
%\phantom{{\it Notes.}} -- In the dwarf irregular bursting model, a single burst is assumed.\\
%\phantom{{\it Notes.}} -- In the dwarf irregular continuous model, the burst duration is of 13 Gyr.
\end{minipage}
\end{center}
\end{table*}
%
%________________________________________________________________

\subsection{Stellar yields}

For all the chemical elements studied, the stellar yields considered in this 
work are the same as the ones used in Paper~I, except for nitrogen. These 
include the yields of \citet{nomoto97a} for massive stars (${\rm M} > 
10$~M$_{\odot}$), the yields of \citet{hoeck97} for low- and intermediate-mass 
stars ($0.8 \leq {\rm M/M}_{\odot} \leq 8$), and the yields of 
\citet{nomoto97b} for Type~Ia SNe (model W7). For Zn we consider the specific 
prescriptions calculated by \citet{calura03} and for Ni we adopt the 
prescriptions presented in Paper~I. 

Our motivation to consider other yields for N comes from the lack of 
reproducing the [N/$\alpha$] ratios of the DLAs studied in Paper~I (see 
Figs.~15 and 17) and from the work by \citet{chiappini03}. This work shows that 
by computing the same chemical evolution models as those used in this paper 
with the new N yields of \citet{meynet02} which take into account the stellar 
rotation, the overestimation of the DLA [N/$\alpha$] ratios is solved. The 
yields with rotation are similar to the ones of \citet{hoeck97} for the 
low-mass stars, but they are substantially smaller for the intermediate-mass 
stars. We hence adopt the same tables of N yields as the ones of 
\citet{chiappini03} extrapolated from the stellar model grids of 
\citet{meynet02}. Meynet \& Maeder did not provide the yields for other 
elements yet. The use of a different set of yields for N and other elements
should, however, not be a problem, because the stellar mass range for the 
production of the other elements considered in this work is different. Indeed, 
O, Si, S, and Mg are mainly produced by massive stars, while N is mainly 
produced by intermediate-mass stars.

Recently, \citet{francois04} published a new set of yields for several elements 
studied in this work, in particular Zn, Ni, O, and Mg. These yields were 
derived from a careful comparison between the Milky Way chemical evolution 
model predictions and the very accurate UVES measurements of abundance ratios 
observed in Galactic stars with metallicities $-4 \leq {\rm [Fe/H]} \leq 0$. 
For the elements Ni and Zn, our adopted yields prescriptions 
\citep[see Paper~I and][ respectively]{calura03} are fully consistent with the 
yields of \citet{francois04} over the DLA metallicity range ($-2.5 < 
{\rm [Fe/H]} < -0.5$). For O, the new yields of \citet{francois04} predict 
lower [O/Fe] ratios than our adopted yields of \citet{nomoto97a} at a given 
metallicity. This leads to a better agreement with the [O/Fe] measurements in 
DLAs as discussed in Sect.~\ref{Q0841-2p476}. The largest discrepancy between 
our adopted yields and the ones of \citet{francois04} concerns Mg. 
\citet{nomoto97a} yields of Mg are known to underestimate the observed [Mg/Fe] 
ratios in Galactic stars, as also shown by \citet{chiappini99}. We tested the 
Fran\c cois et~al. Mg yields, and they usually better reproduce the [Mg/Fe]
ratios measured in DLAs with the spiral models, but tend to overestimate the 
[Mg/Fe] ratios with the dwarf irregular models.

%
%________________________________________________________________

\section{DLA star formation histories}\label{SF-histories}

In Paper~I we have demonstrated that the star formation history and the age of 
a DLA galaxy can be constrained from a detailed comparison of the DLA abundance 
patterns with chemical evolution models. The relative abundances when examined
as a function of a metallicity tracer (e.g. [Fe/H], [Zn/H], or [$\alpha$/H])
allow one to determine {\it the star formation history}, and when examined as a 
function of redshift, they determine {\it the age} of the DLA galaxy defined as 
the time at which the galaxy started to form stars. {\it The star formation 
rate} is also a direct output of the model comparison. In the case of the 
spiral model, the derived SFR is the SFR per unit area and it corresponds to 
the SFR that the galaxy has when it is observed. In the case of the dwarf 
irregular model, to get the SFR per unit area from the SFR output, we assume a 
spherical symmetry and a galactic radius of 3.5 kpc equivalent to the mean size 
of the Magellanic Clouds \citep[]{russell92}\footnote{The radius of dwarf 
galaxies at high redshift is an unknown quantity. In the paper, we assume a
radius of 3.5 kpc. In the case that the radius is half this value, it would 
produce star fomation rates larger by a factor of 4. This difference would, 
however, have no impact on the results described in 
Sect.~\ref{global-picture}.}. It corresponds to the average SFR integrated 
over the time of the burst of star formation, i.e. over the period when the 
star formation is active.
 
The relative abundances that best constrain the chemical evolution model are 
[O/Fe], [Si/Fe], [S/Fe], [S/Zn], [Mg/Fe], [Ni/Fe], [N/O], [N/Si], and [N/S],
most of them accessible for each DLA in our sample (see 
Table~\ref{abundances}). For each DLA individually, we look for the model that 
best reproduces all these abundance ratios as a function of metallicity using 
the same statistical test as in Paper~I and defined as follows. We first 
determine the minimal distance between the data point and the curve 
corresponding to a given chemical model for each abundance ratio. For this 
purpose, we consider the 1~$\sigma$ error on the data point (i.e. the 
covariance matrix of the 1~$\sigma$ measurement error), and we derive the 
minimal distance by computing the distances $d_i$ from the data point to the 
points defining the considered model curve and by determining the $d_i$ for 
which the $d_i$/$\sigma_i$ ratio is minimal. Second, once the minimal distances 
for all the abundance ratios available and for a given model are derived, we 
compute their weighted mean. Finally, the comparison of the weighted means 
obtained for different chemical models determines the best chemical evolution 
model which reproduces the data points and thus the DLA galaxy. The limits are 
not taken into account. We then derive the age of the DLA galaxy by determining 
with the same statistical test at which redshift, the so-called {\it redshift 
of formation}, the deduced best model reproduces all the abundance ratios as a 
function of redshift.

We now describe the deduced star formation histories and ages of the six DLA 
galaxies from our new set of data. In Table~\ref{SFH} we summarize the results. 
We provide the best models reproducing the DLA abundance patterns and the 
corresponding derived redshifts of formation $z_f$, ages, and star formation 
rates per unit area.

%
%________________________________________________________________

\begin{figure*}[t]
\centering
\includegraphics[width=17cm]{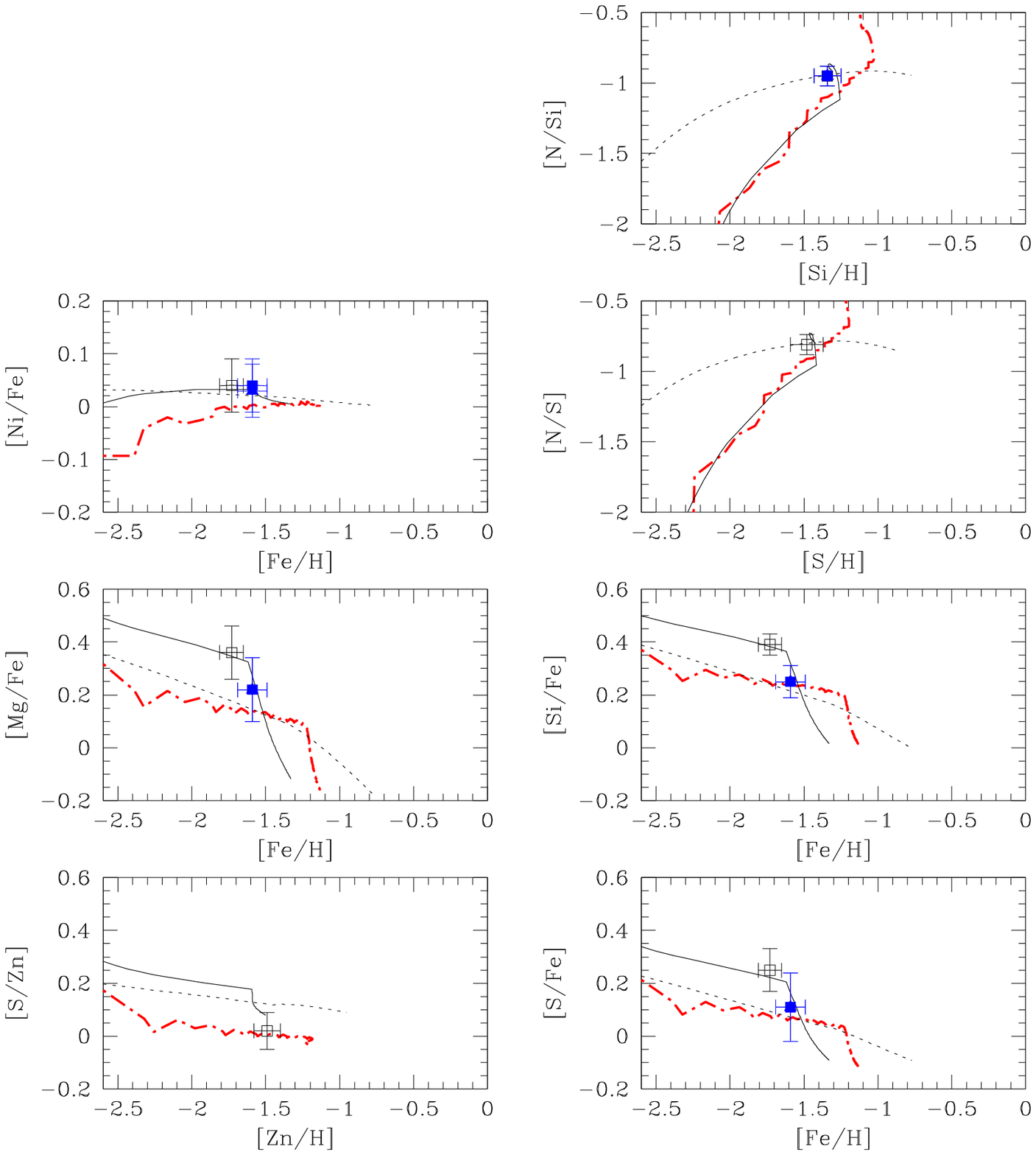}
\caption{Observed and predicted abundance ratios versus metallicity for the DLA
at $z_{\rm DLA} = 2.375$ toward Q\,B0841+129. The thick dashed-dotted curve 
corresponds to the spiral outer disk model, the thin dotted curve to the dwarf 
irregular continuous model with an efficiency $\nu = 0.05$ Gyr$^{-1}$, and the 
thin solid curve to the dwarf irregular bursting model characterized by $\nu = 
0.3$ Gyr$^{-1}$ and a burst duration $\Delta t = 0.2$ Gyr. In this and all the 
following figures, the open squares represent the measured abundance ratios, 
and the filled triangles and the filled squares represent the abundance ratios 
corrected for dust depletion according to the \citet{vladilo02a} method using 
the models E00 and E11, respectively.}
\label{Q0841-2p375-SFH}
\end{figure*}
%
%________________________________________________________________

\subsection{Q\,B0841+129, z$_{\rm DLA}$=2.375}
\label{Q0841-2p375}

The abundance ratios of interest (see above) available in this DLA system are
[Si/Fe], [S/Fe], [S/Zn], [Mg/Fe], [Ni/Fe], [N/Si], and [N/S]. For the refractory
elements we consider the dust-corrected values given in Table~\ref{abundances},
although in this DLA the dust corrections are small, since [Zn/Fe] $= +0.24\pm 
0.02$.

The spiral model which best reproduces the observed abundance patterns is the 
spiral outer disk model (see the thick dashed-dotted curve in 
Fig.~\ref{Q0841-2p375-SFH}). This model fits all the data points within 
1~$\sigma$ error, except the [N/Si] ratio which is at 1.5~$\sigma$. We also 
investigated whether a dwarf irregular model can reproduce the abundance 
patterns. We explored both the dwarf irregular models with a continuous star 
formation regime over the Hubble time characterized by the efficiencies 
$\nu = 0.01$, 0.03, and 0.05 Gyr$^{-1}$, and a bursting star formation regime 
with a burst duration $\Delta t = 0.2$ Gyr and characterized by the 
efficiencies $\nu = 0.1$, 0.3, and 0.5 Gyr$^{-1}$. The bursting model with 
$\nu = 0.3$ Gyr$^{-1}$ reproduces all the data points within 1~$\sigma$ error 
(thin solid curve in Fig.~\ref{Q0841-2p375-SFH}), while the continuous model 
with $\nu = 0.05$ Gyr$^{-1}$ also fits well the DLA abundance patterns, except 
the [S/Zn] ratio which is at 1.4~$\sigma$ (thin dotted curve in 
Fig.~\ref{Q0841-2p375-SFH}). 

We thus have a degeneracy between these three models which all well reproduce 
the abundance patterns observed in this DLA galaxy. However, in terms of 
derived ages and star formation rates, this degeneracy mainly concerns the 
bursting versus continuous star formation regimes. The values given in
Table~\ref{SFH}, indeed, show that the spiral outer disk model and the dwarf 
irregular bursting model, characterized by an episodic bursting and a single 
burst star formation history, respectively (see Sect.~\ref{models}), lead to 
very similar results (age and SFR), while those of the dwarf irregular 
continuous model are different. 
%JXP -- Ad I noted in my email, I don't feel the following is justified.  We
%  can do it, but we must be clear that this is arbitrary and that we will 
%  discuss the degeneracies at greater length in the following section (and that
%  they are summarized in Table 2).
The only way to break this degeneracy is to impose that all the abundance 
patterns have to be within 1~$\sigma$ of the model curve (i.e. have the minimum 
$\chi^2$). The resulting favored model then is the dwarf irregular bursting 
model characterized by $\nu = 0.3$ Gyr$^{-1}$ and $\Delta t = 0.2$ Gyr. 
However, this selection is very arbitrary. 
%In Table~\ref{SFH}, we summarize the properties of the three possible models. 

%
%________________________________________________________________

\begin{figure*}[t]
\centering
\includegraphics[width=17cm]{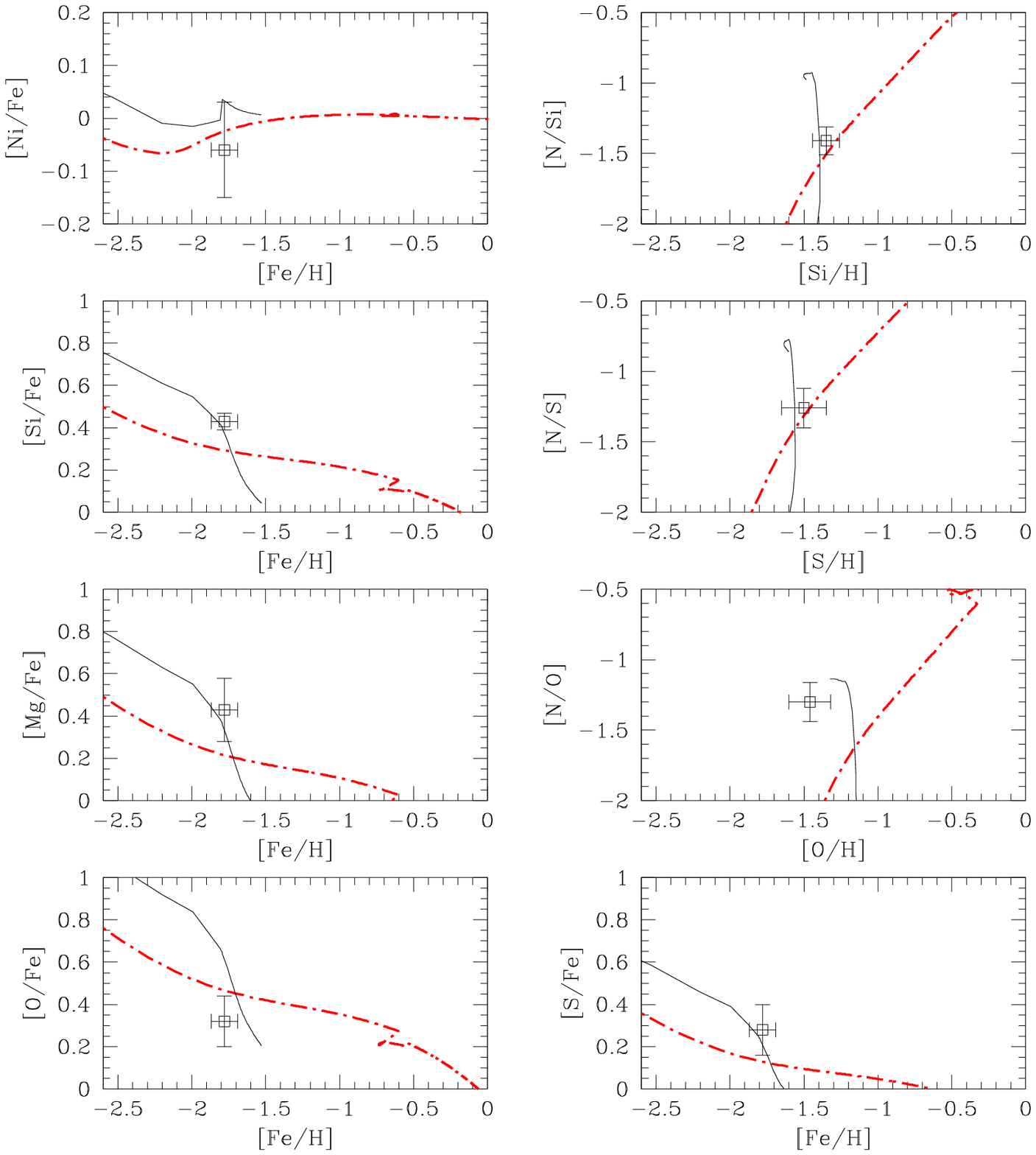}
\caption{Observed and predicted abundance ratios versus metallicity for the DLA
at $z_{\rm DLA} = 2.476$ toward Q\,B0841+129. The thick dashed-dotted curve 
corresponds to the spiral solar neighborhood model and the thin solid curve to 
the dwarf irregular bursting model characterized by an efficiency $\nu = 2$ 
Gyr$^{-1}$ and a burst duration $\Delta t = 0.02$ Gyr. For the definition of 
symbols, see Fig.~\ref{Q0841-2p375-SFH}.}
\label{Q0841-2p476-SFH}
\end{figure*}
%JXP -- The N/Si curve for the dwarf is very strange..
%
%________________________________________________________________

From the analysis of the abundance ratios as a function of redshift, we 
obtain a redshift of formation, $z_f$, of the DLA galaxy between 2.6 and 3 
for the favored model. This corresponds to an age of $0.2-0.6$ Gyr, the DLA 
being observed at $z_{\rm DLA} = 2.375$. The star formation rate per unit area 
is $4\times 10^{-3}$ M$_{\odot}$~yr$^{-1}$~kpc$^{-2}$. %null in 
%this case, because the DLA galaxy is older than the burst duration which means 
%that we observe the galaxy after its burst of star formation, namely during its 
%quiescent phase.

%
%________________________________________________________________

\subsection{Q\,B0841+129, z$_{\rm DLA}$=2.476}
\label{Q0841-2p476}

We have at disposal the following abundance ratios in this DLA system: [Si/Fe], 
[S/Fe], [Mg/Fe], [Ni/Fe], [N/Si], [N/S], and the rare [O/Fe] and [N/O]. With 
[Zn/Fe] $= 0.04\pm 0.11$, the system is likely free from dust and thus 
unaffected by dust depletion effects. No dust corrections have been computed.

%
%________________________________________________________________

\begin{figure*}[t]
\centering
\includegraphics[width=17cm]{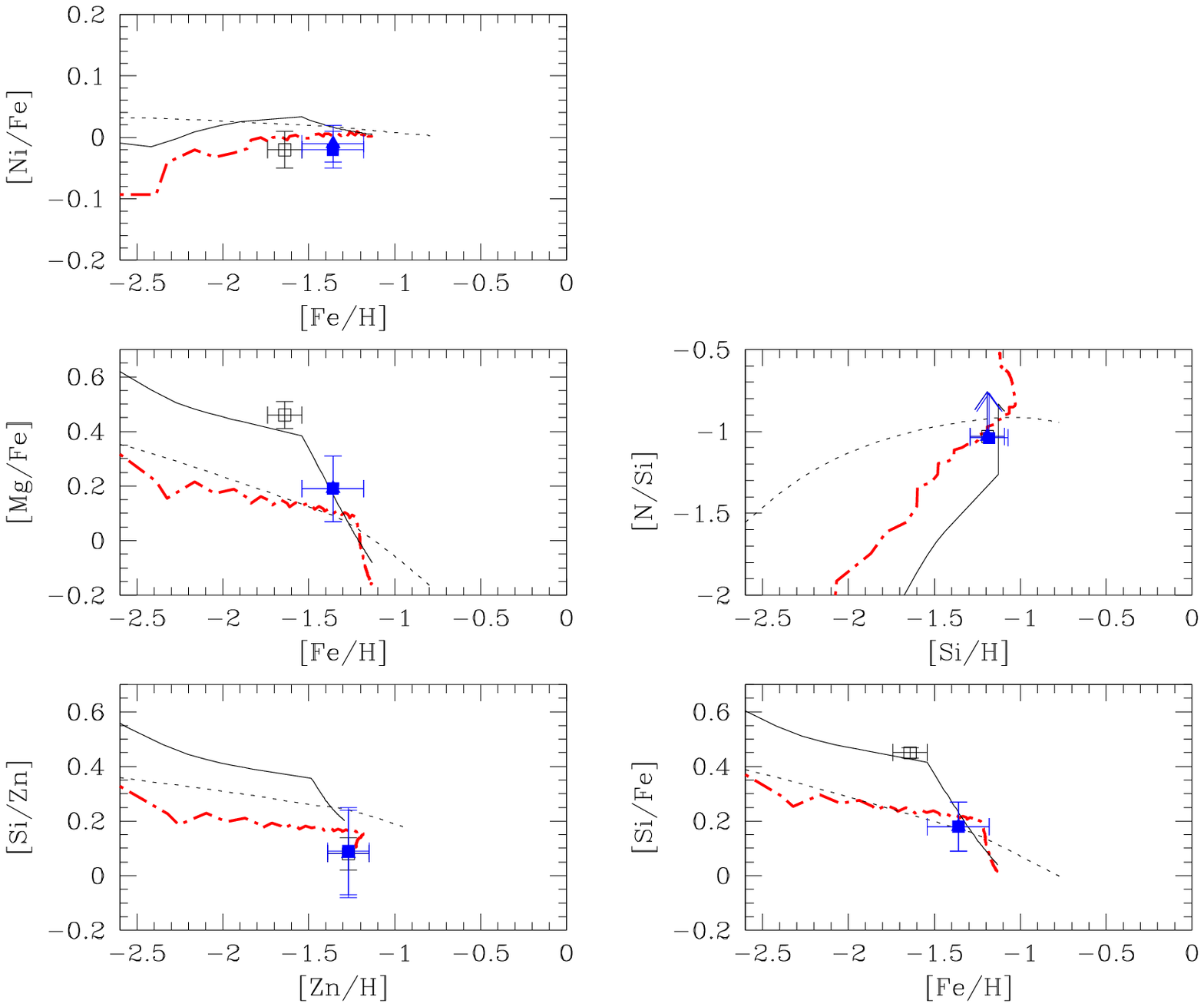}
\caption{Observed and predicted abundance ratios versus metallicity for the DLA
at $z_{\rm DLA} = 1.944$ toward PKS\,1157+014. The thick dashed-dotted curve 
corresponds to the spiral outer disk model, the thin dotted curve to the dwarf 
irregular continuous model having an efficiency $\nu = 0.05$ Gyr$^{-1}$, and 
the thin solid curve to the dwarf irregular bursting model characterized by an 
efficiency $\nu = 0.7$ Gyr$^{-1}$ and a burst duration $\Delta t = 0.1$ Gyr. 
For the definition of symbols, see Fig.~\ref{Q0841-2p375-SFH}.}
\label{Q1157-SFH}
\end{figure*}
%JXP -- The N/Si curve for the spiral is odd.  
%
%________________________________________________________________

None of the spiral models succeeds in reproducing the [$\alpha$/Fe] ratios in 
this DLA galaxy. Indeed, the [O/Fe] ratio is located at 1.2~$\sigma$, [Si/Fe] 
at 2.4~$\sigma$, [S/Fe] at 1.3~$\sigma$, and [Mg/Fe] at 1.4~$\sigma$ from the 
best spiral model curve (see the thick dashed-dotted curve in 
Fig.~\ref{Q0841-2p476-SFH}). Moreover, the [N/O] ratio is also not fitted 
(3.1~$\sigma$). Similarly, none of dwarf irregular continuous models fits the 
observed abundance patterns. We investigated dwarf irregular bursting models 
with efficiencies between 1.5 to 4 Gyr$^{-1}$ by step of 0.5 Gyr$^{-1}$ and 
burst durations between 0.01 and 0.05 Gyr. The model with $\nu = 2$ Gyr$^{-1}$ 
and a very short burst duration $\Delta t = 0.02$ Gyr best reproduces the 
observed abundance patterns. Indeed, the model fits all the data points within 
1~$\sigma$ error, except the [O/Fe] and [N/O] ratios (thin solid curve in 
Fig.~\ref{Q0841-2p476-SFH}). 

O is the most reliable diagnostic in the SFH studies, thus the fact that the
[O/Fe] and [N/O] ratios are not well reproduced by the model may suggest either
that the O abundance measurement is underestimated due to a possible \ion{O}{i} 
line saturation (see Paper~II) or that the adopted O yields are not reliable
enough. When considering the new yields of \citet{francois04} which predict a 
lower O abundance at a given metallicity, the problem of O is solved for both 
the dwarf irregular bursting model and the best spiral model. %Similarly, the 
%fit of the [Mg/Fe] ratio is improved for the spiral model with the new Mg 
%yields. This is a very positive result, since the old Mg yields are known to
%underestimate the [Mg/Fe] ratio in spirals \citep{samland97,chiappini99}.
%
%JXP -- It seems that if we ignore O, then the spiral model works fine.  Is
%  that right?  If so, I suggest we can't rule it out at all.

The degeneracy between the three star formation histories investigated is in 
this case alleviated. Only the dwarf irregular bursting model allows to 
reproduce the abundance patterns of this DLA characterized by a high 
$\alpha$-enhancement. We would like to point out that the adopted Salpeter IMF 
in the dwarf irregular models largely contributes to the successful fit of 
these DLA data points. Indeed, a more top-heavy IMF relative to the Scalo IMF 
is needed to reproduce the high $\alpha$/Fe ratios. This is why the spiral 
models are ruled out.
%JXP -- I wonder, do we find these kind of alpha/Fe enhancements in dwarf galaxies
%  today?  I thought many (most?) have roughly solar alpha/Fe.
%  And, again, I think it is O that rules out the sprials, not the IMF (aside from
%  how it pertains to O).

%
%________________________________________________________________

\begin{figure*}[t]
\centering
\includegraphics[width=17cm]{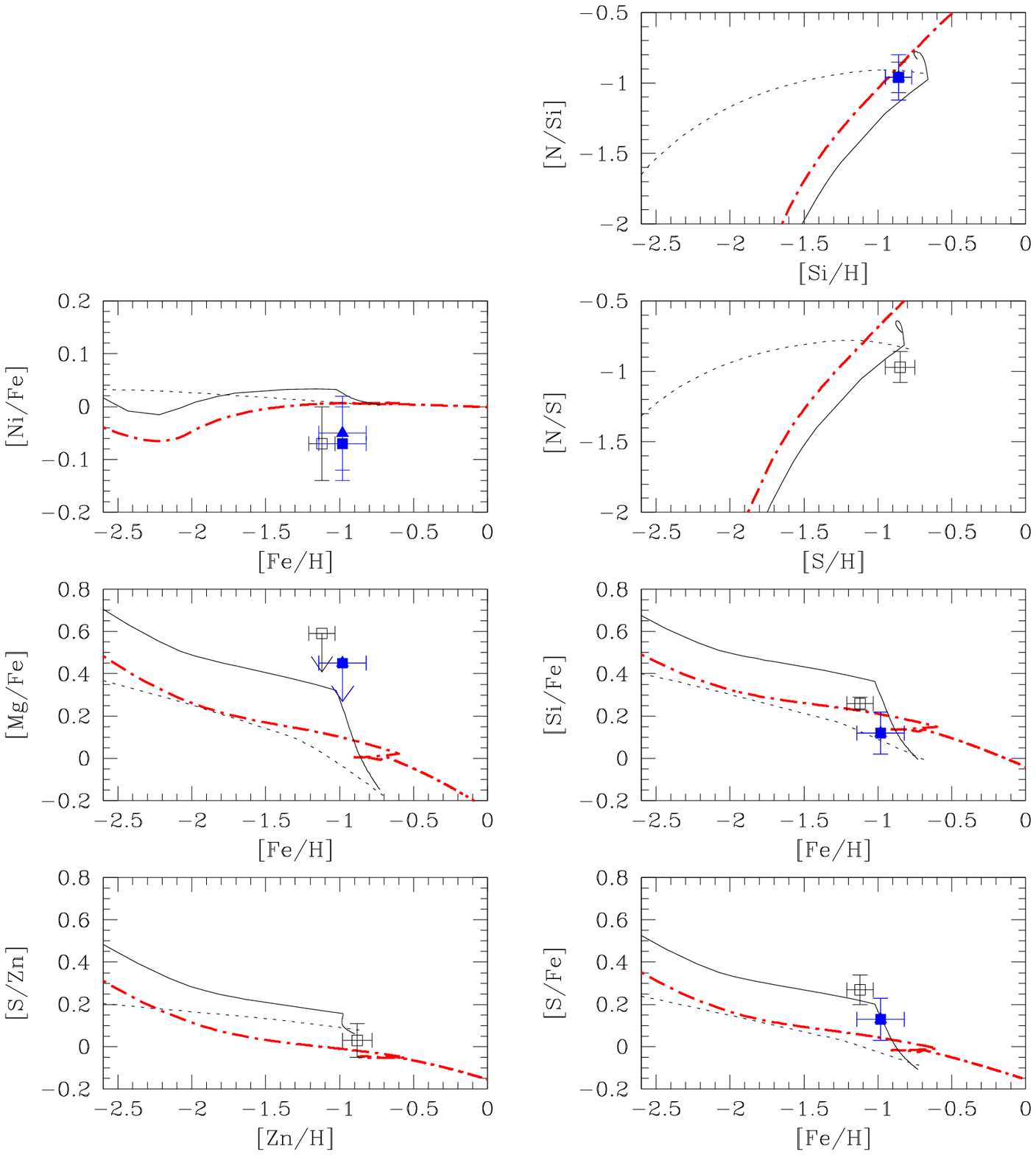}
\caption{Observed and predicted abundance ratios versus metallicity for the DLA
at $z_{\rm DLA} = 1.892$ toward Q\,B1210+175. The thick dashed-dotted curve 
corresponds to the spiral inner disk model, the thin dotted curve to the dwarf 
irregular continuous model with an efficiency $\nu = 0.06$ Gyr$^{-1}$, and the 
thin solid curve to the dwarf irregular bursting model characterized by an 
efficiency $\nu = 1.2$ Gyr$^{-1}$ and a burst duration $\Delta t = 0.2$ Gyr. 
For the definition of symbols, see Fig.~\ref{Q0841-2p375-SFH}.}
\label{Q1210-SFH}
\end{figure*}
%
%________________________________________________________________

From the abundance ratios versus redshift diagrams, we find a redshift of 
formation $z_f \simeq 2.5$ for this DLA galaxy, while we observe this DLA at 
$z_{\rm DLA} = 2.476$. This corresponds to a very young age of $0.02\pm 0.01$ 
Gyr only due to the very short burst duration, and suggests that the DLA galaxy 
is likely experiencing its first burst of star formation. The measured star 
formation rate per unit area is $3\times 10^{-2}$ 
M$_{\odot}$~yr$^{-1}$~kpc$^{-2}$. Examples of galaxies with such a short burst 
duration exist among the local metal-poor galaxies. In particular, 
\citet{kunth95} studied the chemical properties of IZw18 by means of a 
starburst model similar to ours, characterized by a burst duration of 0.01 Gyr 
and an average SFR of $\sim 0.01$ M$_{\odot}$~yr$^{-1}$~kpc$^{-2}$. Transient 
episodes of star formation with such a duration are also likely to be typical 
of some local blue irregular galaxies \citep{annibali03}.

%JXP -- This is beyond the scope of this paper, but I have started giving
% serious thought to mixing.  Consider, for example, the case of Q0841+129, z=2.476
% which is characterized by such a young Dt=20Myr.  I wonder if it makes any
% sense to argue that we'd ever see the signature of such a SFH.  That is, 
% we are ignoring the real 'gap' that exists between where the metals are 
% formed (stars) and where we observe them (the ambient ISM).  I think we 
% will agree that the DLA gas does not generally probe the inner parts of a 
% SF region.  Therefore, the metals we do observe need to be transported (e.g.
% via SN winds or turbulence) significantly beyond the SF region.  I would doubt
% this transport could be accomplished in ~20Myr.  So the gas we are seeing is
% almost certainly older than this.  So, I wonder how we could ever really see the
% signature of such a young burst.  In my mind it would either require (i) all SF
% shut off after 20Myr [but how do you stop the IMS and TypeIa SN from contribution;
% or (ii) we are seeing the SN products being carried by a SN wind directly [not
% sure this makes sense either].  Anyhow, it is fair to say that the observed 
% abundances can be modeled by such a short burst of SF, but I really wonder
% how this all works with mixing, etc.
%  Again, this is beyond the scope of this paper, but it is something to
%  start thinking about.  I suspect we are going to be worrying a lot about 
%  mixing in the next decade.

%
%________________________________________________________________

\subsection{PKS\,1157+014, z$_{\rm DLA}$=1.944}
\label{Q1157}

Few abundance ratios of interest are available in this DLA system: [Si/Fe], 
[Si/Zn], [Mg/Fe], [Ni/Fe], and a lower limit to [N/Si]. We consider the 
dust-corrected values given in Table~\ref{abundances}, since they are 
relatively important in this system with [Zn/Fe] $= +0.37\pm 0.07$.

%
%________________________________________________________________

\begin{figure*}[t]
\centering
\includegraphics[width=17cm]{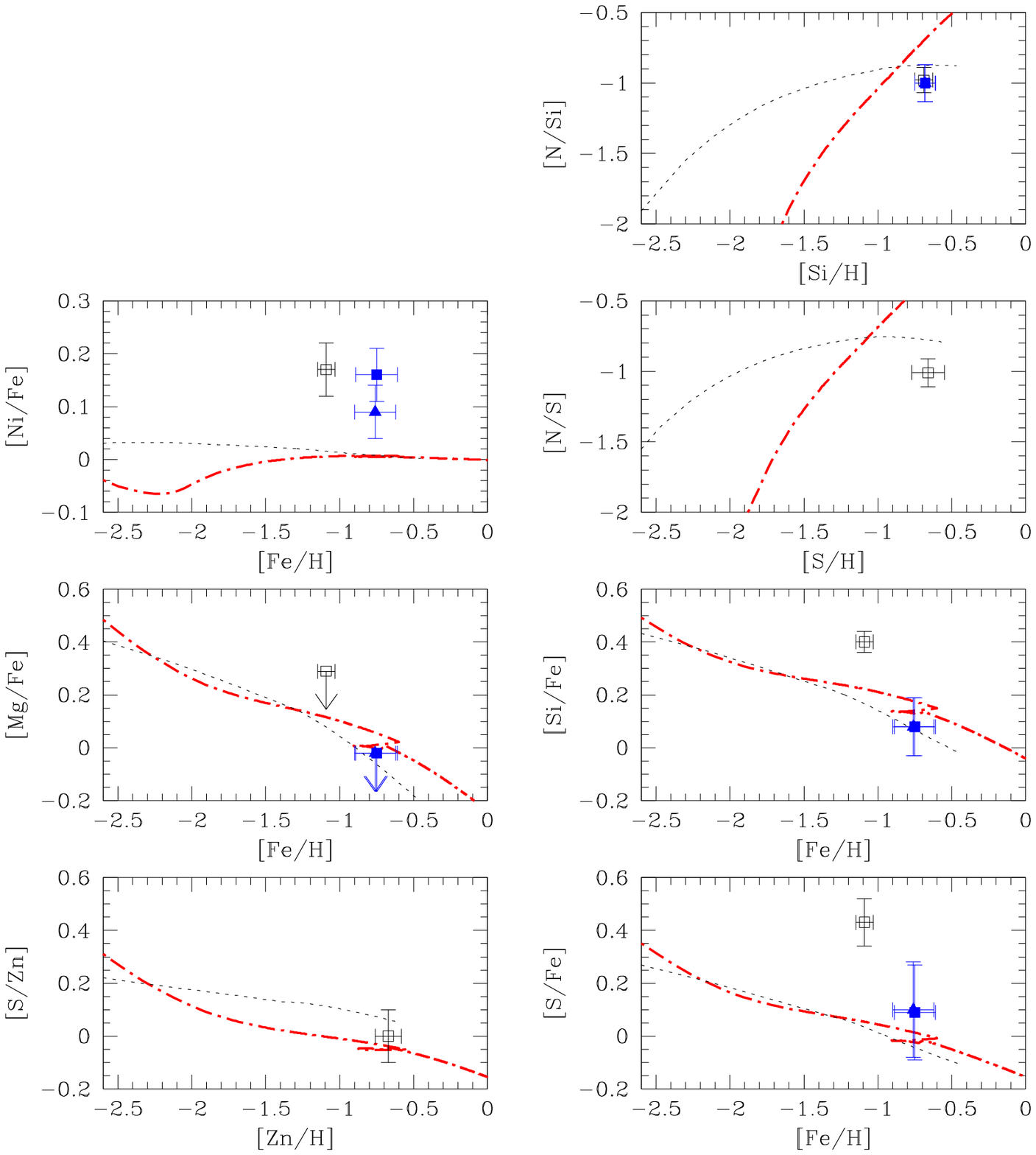}
\caption{Observed and predicted abundance ratios versus metallicity for the DLA
at $z_{\rm DLA} = 1.864$ toward Q\,B2230+02. The thick dashed-dotted curve 
corresponds to the spiral inner disk model and the thin dotted curve to the 
dwarf irregular continuous model with an efficiency $\nu = 0.1$ Gyr$^{-1}$. For 
the definition of symbols, see Fig.~\ref{Q0841-2p375-SFH}.}
\label{Q2230-SFH}
\end{figure*}
%
%________________________________________________________________

The best spiral model is the spiral outer disk model. It reproduces all the 
data points within 1~$\sigma$ error and is in agreement with the lower limit to 
[N/Si] (see the thick dashed-dotted curve in Fig.~\ref{Q1157-SFH}). We 
considered also dwarf irregular models with both the continuous and bursting 
star formation regimes. For both of them we can constrain adequately the star
formation parameters to reproduce successfully all the abundance patterns 
within 1~$\sigma$ error: for the continuous model an efficiency $\nu = 0.05$ 
Gyr$^{-1}$ is needed (thin dotted curve in Fig.~\ref{Q1157-SFH}), and for the 
bursting model a burst with an efficiency $\nu = 0.7$ Gyr$^{-1}$ and a duration 
$\Delta t = 0.1$ Gyr is needed (thin solid curve in Fig.~\ref{Q1157-SFH}). This 
leads again to an important degeneracy between these three models tracing three
different star formation histories: an episodic bursting SFH, a continuous SFH,
and a single burst SFH, respectively. If we compare the weighted means of the 
minimal distances between the model curves and the available data points, then 
the episodic bursting SFH (spiral outer disk model) with the minimum $\chi^2$ 
is the favorite one. However, there is no physical reason to exclude the two 
other star formation histories, since they all agree with the abundance 
patterns. 
%
%In addition, a problem of timescale is observed between [Si/Zn] and the other 
%[$\alpha$/Fe] ratios for the spiral model and the bursting dwarf irregular 
%model, as in the DLA at $z_{\rm DLA} = 2.375$ toward Q0841+129 (see 
%Sect.~\ref{Q0841-2p375}). This could suggest that the best model is at the end 
%the continuous dwarf irregular model, but the problem is solved when 
%considering the new Zn yields of \citet{francois04}. However, the [Mg/Fe] 
%ratio is in this case highly overestimated with the Mg yields of 
%\citet{francois04}.

%
%________________________________________________________________

\begin{figure*}[t]
\centering
\includegraphics[width=17cm]{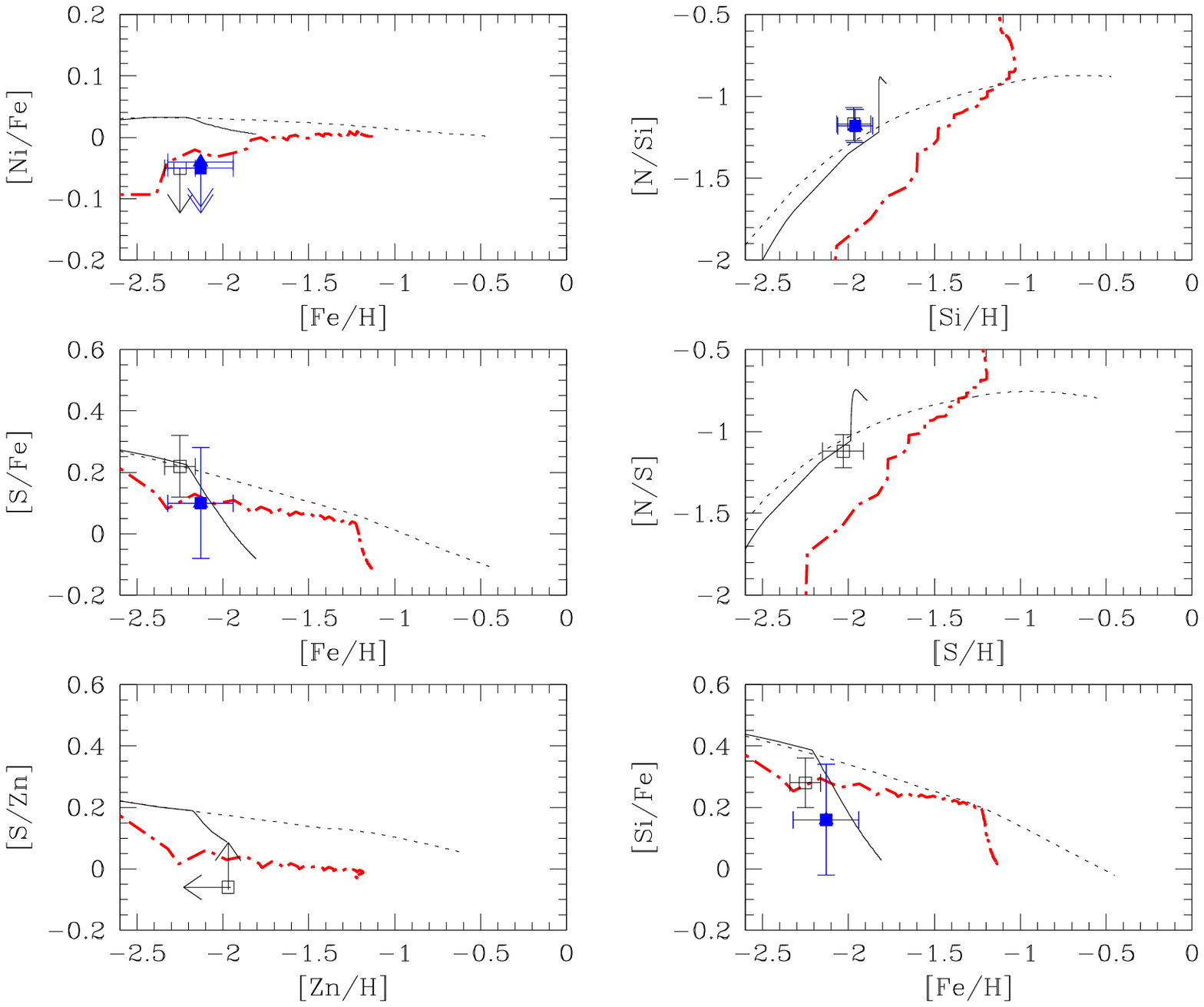}
\caption{Observed and predicted abundance ratios versus metallicity for the DLA
at $z_{\rm DLA} = 2.279$ toward Q\,B2348$-$1444. The thick dashed-dotted curve 
corresponds to the spiral outer disk model, the thin dotted curve to the dwarf 
irregular continuous model with an efficiency $\nu = 0.1$ Gyr$^{-1}$, and the 
thin solid curve to the dwarf irregular bursting model characterized by an 
efficiency $\nu = 0.1$ Gyr$^{-1}$ and a burst duration $\Delta t = 0.15$ Gyr. 
For the definition of symbols, see Fig.~\ref{Q0841-2p375-SFH}.}
\label{Q2348-SFH}
\end{figure*}
%
%________________________________________________________________

From the abundance ratios studied as a function of redshift, we derived the 
redshift of formation and the age of the DLA galaxy for the three best models. 
The obtained age proves to be very different from one model to the other. 
Indeed, for the spiral outer disk model we find $z_f = 2-2.4$ which corresponds 
to an age of $0.1-0.6$ Gyr, for the dwarf irregular bursting model we have $z_f 
= 2.4-2.9$ which gives an intermediate age of $0.7-1.2$ Gyr, and for the dwarf 
irregular continuous model we get $z_f = 4-5.5$ which corresponds to a much 
older age of $1.9-2.5$ Gyr. The star formation rate per unit area is also very
disparate from one model to the other: $7.5\times 10^{-3}$ 
M$_{\odot}$~yr$^{-1}$~kpc$^{-2}$ for the spiral model, between $7.5\times 
10^{-4}$ and $1\times 10^{-3}$ M$_{\odot}$~yr$^{-1}$~kpc$^{-2}$ for the dwarf 
irregular continuous model, and $1.6\times 10^{-2}$ 
M$_{\odot}$~yr$^{-1}$~kpc$^{-2}$ for the dwarf irregular bursting model.

%
%________________________________________________________________

\subsection{Q\,B1210+175, z$_{\rm DLA}$=1.892}
\label{Q1210}

We have at disposal the following abundance ratios in this DLA system: [Si/Fe],
[S/Fe], [S/Zn], [Ni/Fe], [N/Si], [N/S], and an upper limit to [Mg/Fe]. We
consider the dust-corrected values given in Table~\ref{abundances} for the 
abundance ratios including a refractory element, although the dust content in 
this system is relatively low, [Zn/Fe] $= +0.24\pm 0.07$.

The spiral model which best reproduces the abundance patterns is the spiral
inner disk model. This model fits all the data points within 1~$\sigma$ error, 
except the [N/S] ratio which is at 2.9~$\sigma$ (see the thick dashed-dotted 
curve in Fig.~\ref{Q1210-SFH}). We investigated also the dwarf irregular models 
first to see whether the [N/S] ratio can better be reproduced and second 
because the probability that the QSO line of sight intercepts the internal 
parts of a spiral galaxy is low. None of the dwarf irregular models with a 
continuous star formation regime can reproduce the [N/S] ratio within 
1~$\sigma$ error. The best of these models is the model with an efficiency 
$\nu = 0.06$ Gyr$^{-1}$, but it fails to fit the [S/Fe] ratio at 1.4~$\sigma$ 
in addition to the [N/S] ratio which is at 1.3~$\sigma$ (thin dotted curve in 
Fig.~\ref{Q1210-SFH}). The favored model is the dwarf irregular bursting model 
with an efficiency $\nu = 1.2$ Gyr$^{-1}$ and a duration $\Delta t = 0.2$ Gyr 
(thin solid curve in Fig.~\ref{Q1210-SFH}). Indeed, this model reproduces all 
the abundance patterns within 1~$\sigma$ error and is in agreement with the 
upper limit to [Mg/Fe] as well as the lower limit to [Mg/Fe] obtained from the 
\ion{Mg}{ii}\,$\lambda$2796,2803 doublet which dust-corrected value is 
[Mg/Fe]$_{\rm cor} > +0.37$ (see Paper~II).
%
%However, we observe an incompatibility in the timescale between [S/Zn] and the 
%other abundance ratios for this model. This is nevertheless once again solved 
%when considering the new Zn yields of \citet{francois04}. 

The analysis of the abundance ratios as a function of redshift gives for the
favored model a redshift of formation $z_f = 2.2-2.6$ of this DLA galaxy. This
corresponds to an age of $0.5-1$ Gyr. The star formation rate is $1.7\times
10^{-2}$ M$_{\odot}$~yr$^{-1}$~kpc$^{-2}$.
%null here again, because the DLA is older than the burst duration and hence we 
%observe the galaxy after its burst of star formation.

%
%________________________________________________________________

\subsection{Q\,B2230+02, z$_{\rm DLA}$=1.864}
\label{Q2230}

This DLA galaxy has a very high metallicity, [Zn/H] $= -0.67\pm 0.09$, higher 
than 1/5 solar, and relatively important dust depletion corrections that may
bring an uncertainty in the study of the star formation history of this object 
(see Table~\ref{abundances}). The abundance ratios at disposal are [Si/Fe], 
[S/Fe], [S/Zn], [Ni/Fe], [N/Si], [N/S], and an upper limit to [Mg/Fe].

Only the spiral inner disk model manages to reproduce the observed high 
metallicity of this DLA galaxy. However, this model fails to fit the [N/Si] and 
[N/S] ratios which are at 2 and 4.5~$\sigma$ from the model curve, 
respectively, and the [Mg/Fe] upper limit (see the thick dashed-dotted curve in 
Fig.~\ref{Q2230-SFH}). We further investigated continuous star formation 
histories, but with weaker star formation efficiencies considering $\nu$ 
parameters between 0.05 and 0.2 Gyr$^{-1}$. We found that the dwarf irregular 
continuous model with $\nu = 0.1$ Gyr$^{-1}$ best reproduces the abundance 
patterns of this DLA galaxy (thin dotted curve in Fig.~\ref{Q2230-SFH}). It is 
in agreement with the upper limit to [Mg/Fe] and fits all the data points 
within 1~$\sigma$ error, except the [N/S] ratio at 2.3~$\sigma$. The oversolar 
[Ni/Fe] measurement can hardly be explained by any chemical evolution model, 
since Ni should closely trace Fe. 
%The \ion{Ni}{ii} column density measurement seems to be, however, very accurate 
%(see Paper~II). 

%
%When considering the new Ni yields of \citet{francois04}, the predicted [Ni/Fe] 
%ratio is, on the contrary, far too oversolar and does not agree with any 
%[Ni/Fe] ratio measured in any DLA. The new yields of Zn prove to be, on the
%other hand, well improved and in a very good agreement with the DLA abundance
%measurements, since they help to solve the problem of the different timescales 
%predicted by [S/Zn] and the other abundance ratios encountered when using the 
%old Zn yields and this for all the DLAs studied. The new Mg yields lead in this
%case to a too high [Mg/Fe] ratio compared to the observed upper limit. As
%mentioned in Sect.~\ref{Q0841-2p476}, the [Mg/Fe] is known to be underestimated
%by our adopted yields, but the Mg yields of \citet{francois04} tend to 
%overestimate the [Mg/Fe] ratios relatively to those measured in DLAs, as also 
%shown in Sect.~\ref{Q1157}.
%

We used the abundance ratios versus redshift to estimate the redshift of
formation and the age of this DLA galaxy for the favored model identified. The 
redshift of formation found is very high, $z_f > 10$, implying a very old age,
higher than 3.2 Gyr, for this galaxy. This suggests that galaxies were already
formed at $z > 10$. The star formation rate per unit area of this DLA galaxy is 
of $6.5\times 10^{-4}$ M$_{\odot}$~yr$^{-1}$~kpc$^{-2}$.

%
%________________________________________________________________

\subsection{Q\,B2348--1444, z$_{\rm DLA}$=2.279}
\label{Q2348}

This DLA, on the contrary to the DLA toward Q\,B2230+02, has a low metallicity
1/180 solar ([Fe/H] $= -2.25\pm 0.09$). The derived upper limit to the Zn 
abundance indicates weak effects due to dust depletion, since [Zn/Fe] 
$< +0.28$. The abundance ratios at disposal for the study of the star formation 
history of this DLA galaxy are [Si/Fe], [S/Fe], [N/Si], [N/S], and the upper 
limits to [Ni/Fe] and [S/Zn].

%
%________________________________________________________________

\begin{figure*}[t]
\centering
\includegraphics[width=18cm]{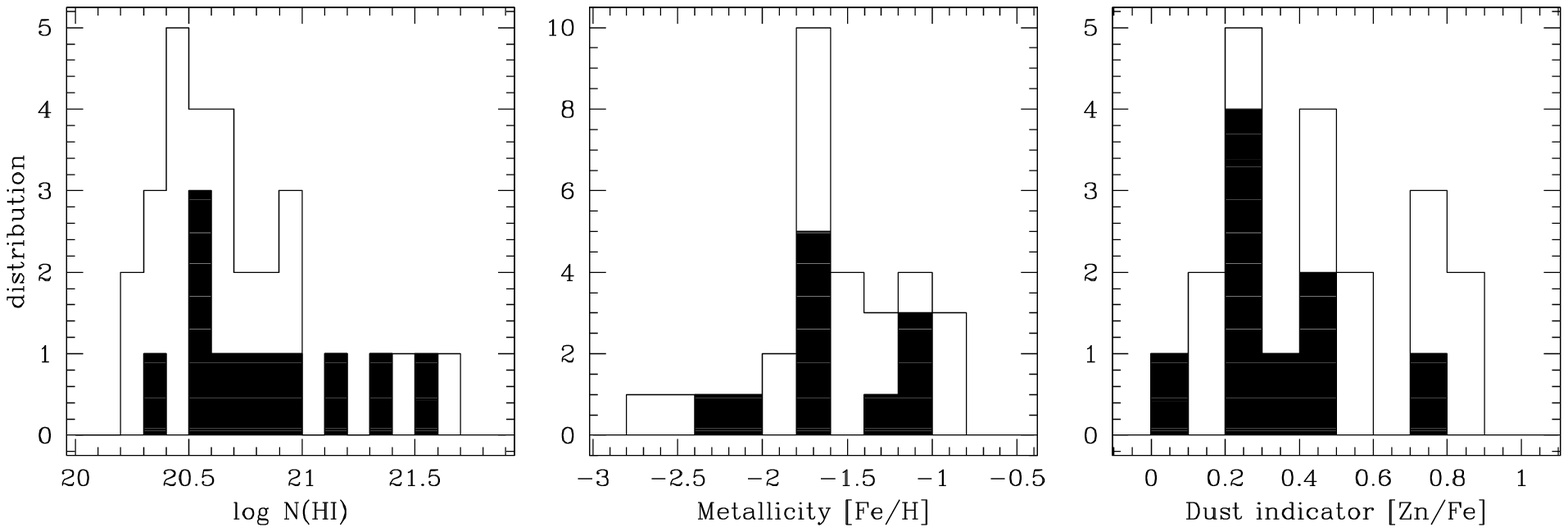}
\caption{{\it From the left to the right panel.} Histograms of the \ion{H}{i} 
column density, metallicity [Fe/H], and dust depletion indicator [Zn/Fe] 
distributions for the whole sample of known DLAs with metallicity measurements 
in the redshift interval $z_{\rm DLA} = 1.7-2.5$ and our sample of eleven DLAs 
studied in Papers~I and II (filled area). The comparison of these distributions 
for the two sets of data shows that the DLAs in our sample cover the full range 
of values for these three properties as observed in DLAs.}
\label{histograms}
\end{figure*}
%form a relatively unbiased sample. It only appears to be slightly biased 
%toward high \ion{H}{i} column densities.
%JXP -- We need to reference the source of the 'control' sample.
%
%________________________________________________________________

The spiral model which best reproduces the abundance patterns is the spiral 
outer disk model. This model fits very well the [$\alpha$/Fe] ratios, but fails 
to reproduce the [N/Si] and [N/S] ratios which are at 4.0~$\sigma$ and 
2.3~$\sigma$, respectively (see the thick dashed-dotted curve in 
Fig.~\ref{Q2348-SFH}). We then considered dwarf irregular models with both the 
continuous and bursting star formation regimes to see whether they better 
reproduce the data points. The best dwarf irregular continuous model is 
characterized by an efficiency $\nu = 0.1$ Gyr$^{-1}$ (see the thin dotted 
curve in Fig.~\ref{Q2348-SFH}). It reproduces all the abundance patterns within 
1~$\sigma$, except the [Si/Fe] ratio which is at 1.1~$\sigma$. Similarly, the 
dwarf irregular bursting model, characterized by an efficiency $\nu = 0.1$ 
Gyr$^{-1}$ and a burst duration $\Delta t = 0.15$ Gyr, also reproduces all the 
data points within 1~$\sigma$, except the [N/Si] ratio which is at 1.4~$\sigma$ 
(see the thin solid curve in Fig.~\ref{Q2348-SFH}). 
%However, none of these models agrees with the [Ni/Fe] upper limit which implies 
%a far too low [Ni/Fe] ratio relatively to the model predictions. 
We thus have for this DLA galaxy a high degeneracy between these two dwarf 
irregular models tracing two different star formation histories. 

From the abundance ratios studied as a function of redshift, we derived the
redshift of formation and the age of the DLA galaxy for the two possible SFHs. 
For the dwarf irregular continuous model we find $z_f = 2.6-2.7$, while for the 
dwarf irregular bursting model we find $z_f = 2.4-2.5$. This corresponds to 
ages of relatively the same order, $0.3-0.5$ Gyr and $0.1-0.3$ Gyr, 
respectively. The star formation rates per unit area obtained for the two 
models are, on the contrary, relatively different, $8.7\times 10^{-4}$ and 
$2.2\times 10^{-3}$ M$_{\odot}$~yr$^{-1}$~kpc$^{-2}$, respectively.

%
%________________________________________________________________

\section{Comparison of DLA properties with other high-redshift galaxies}
\label{global-picture}

The eleven DLA systems at $z_{\rm DLA} = 1.7-2.5$ studied in Papers~I and II 
were selected only on the basis of bright background quasars and the existence 
of HIRES/Keck spectra. They form a sample which covers the full range of 
\ion{H}{i} column densities, metallicities, and dust depletion levels observed 
in other DLAs at similar redshifts, as illustrated in 
Fig.~\ref{histograms}\footnote{The ``control'' sample is a compilation of 
abundance measurements from the literature of all DLAs at $z_{\rm DLA} = 
1.7-2.5$ obtained from high-resolution spectra.}. However, this sample appears 
to be slightly biased toward high \ion{H}{i} column densities, since it 
contains three out of the five higher \ion{H}{i} column density systems of our 
control sample, and toward less dusty DLA galaxies, as indirectly selected with 
the bright QSOs. Nevertheless, these effects are not determinant on our 
results, since they are not significant,  the full range of DLA values is 
covered by the DLAs in our sample. The derived star formation histories of 
nine of these DLA galaxies\footnote{Two DLAs in our sample have high ionization 
fractions (see Sect.~\ref{ionization}). We exclude them from our star formation 
history study, because the measure of their intrinsic abundances is very 
uncertain.} thus are well representative of the average DLA galaxy population.
%{\bf The selection of bright QSOs may introduce a bias against very dusty 
%regions in high-redshift galaxies. They yet seem to form an unbiased sample of 
%DLAs. Indeed, they have \ion{H}{i} column densities, metallicities, dust 
%depletion levels, and abundance patterns which cover the full range of values 
%observed in other DLAs at similar redshifts, as illustrated in 
%Fig.~\ref{histograms}\footnote{The ``control'' sample is a compilation of 
%abundance measurements from the literature of all DLAs at $z_{\rm DLA} = 
%1.7-2.5$ obtained from high-resolution spectra.}. Our sample of DLAs only 
%appears to be slightly biased toward high \ion{H}{i} column densities, since 
%it contains three out of the five higher \ion{H}{i} column density systems of 
%our control sample.} However, ...
 
Our results show that the abundance patterns observed in DLAs are compatible 
with various star formation histories: episodic bursting, single burst, and 
continuous star formation histories with different star formation efficiencies. 
These are typical of inner and outer regions of Milky Way type spiral galaxies 
and of dwarf irregular galaxies with bursting and continuous star formation 
regimes. The degeneracy between these SFHs is unfortunately large for most 
DLAs. It is only broken for DLAs with specific abundance patterns, namely with 
high $\alpha$-enhancements, high metallicities, and/or low N/$\alpha$ ratios. 
This is the case of three DLAs among the nine in our sample. For the six DLAs 
with degenerated SFHs, we assume the favored SFH is the one for which the 
related model reproduces the abundance patterns with the minimum $\chi^2$ (see 
Paper~I and Table~\ref{SFH}). We can, however, note that the main results 
discussed in the following two sub-sections hold despite the degeneracy.
%whatever we choose for the favored SFH.
%JXP -- Again, this is too dangerous a practice.

Let's place the derived DLA properties in the global context of high-redshift 
galaxies and let's compare them with the properties of emission-selected 
galaxies.

%Independently of this degeneracy problem, the general results which comes out 
%from this study are: (i)~the DLA galaxies seem to have star formation histories 
%characterized by {\it low star formation efficiencies}; (ii)~as a consequence, 
%the star formation rates per unit area are {\it moderate} in DLA galaxies, with 
%values between $-3.2 < \log {\rm SFR} < -1.1$ M$_{\odot}$~yr$^{-1}$~kpc$^{-2}$;
%and (iii)~the DLA galaxies show a range of ages with some DLAs being very young 
%with ages between $20-600$ Myr likely experiencing their first star formation 
%episodes and other old with ages longer than 1 Gyr.
%JXP -- Are we saying we can't tell if the gas is 20Myr or several Gyr or that
%  the DLA show a range of ages.  I suspect it is both but it mainly reads that
%  we can't tell what the age is at all (or its true spread).

%
%________________________________________________________________

\begin{figure*}[t]
\centering
\includegraphics[width=17cm]{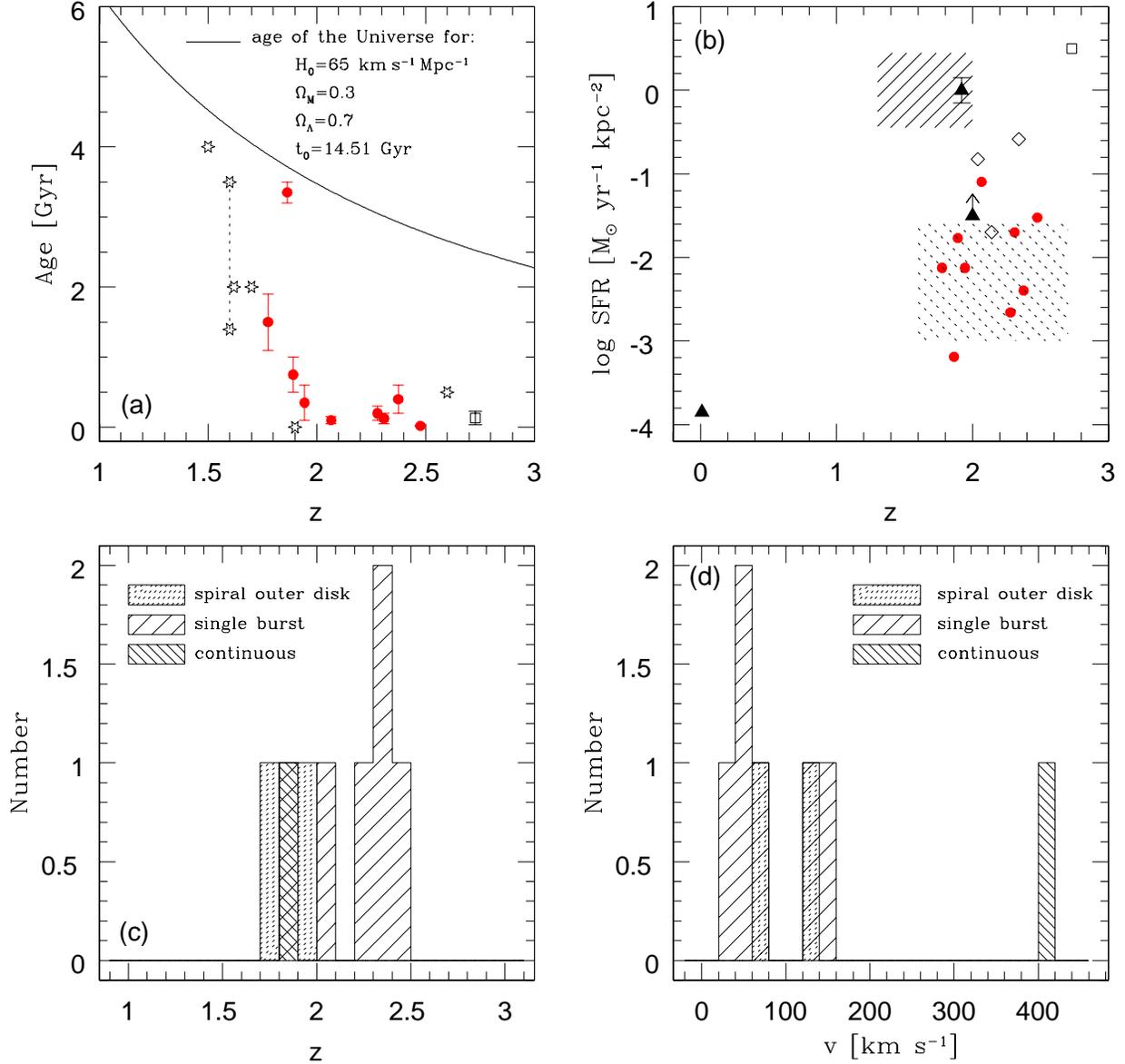}
\caption{{\it Panel}~(a). Age distribution as a function of redshift for the
nine DLA galaxies in our sample (filled circles), for the emission-selected 
galaxies studied by \citet[][ stars]{bruzual02}, and for the Lyman-break galaxy 
MS\,1512$-$c58 \citep[square; ][]{pettini00b,pettini02}. The solid curve 
corresponds to the age limit of the Universe defined by the adopted cosmology. 
{\it Panel}~(b). Star formation rate per unit area distribution as a function 
of redshift. Our DLA results are shown by filled circles. The dotted-shaded 
area corresponds to the SFRs obtained for the DLAs by \citet{wolfe03a} from the 
\ion{C}{ii}$^*$\,$\lambda$1335 measurements and the dashed-shaded area to the 
SFRs obtained for the emission-selected galaxies from the Gemini Deep Deep 
Survey \citep{savaglio04}. The triangles represent the three DLAs for which we 
have an estimation of their SFR from emission lines (from top to bottom we have 
the DLAs toward Q\,B2206$-$1958, PKS\,0458$-$02, and HS\,1543+5921, 
respectively). The square corresponds to the LBG MS\,1512$-$c58 and the 
diamonds to the GRB host galaxies (GRB\,000926, GRB\,011211, and GRB\,021004). 
{\it Panels}~(c) {\it and} (d). Distributions of the star formation histories 
determined for the DLA galaxies studied as a function of redshift and velocity 
width of the low-ionization absorption lines related to the mass of the galaxy,
respectively. Three different SFHs, referred as spiral outer disk (episodic 
bursting SFH), dwarf irregular bursting star formation (single burst SFH), and 
dwarf irregular continuous star formation (continuous SFH), were identified as 
reproducing the DLA abundance patterns.}
\label{age-SFR}
\end{figure*}
%
%________________________________________________________________

\subsection{Age distribution}\label{distr-age}

In Fig.~\ref{age-SFR}\,(a) we show the age distribution as a function of 
redshift for our nine DLA galaxies (circles), for six emission-selected 
galaxies (stars) which age was derived from the analysis of their integrated 
spectra \citep{bruzual02}, and for the Lyman-break galaxy (LBG) MS\,1512$-$cB58
\citep[square;][]{pettini00b,pettini02}. MS\,1512$-$cB58 has age properties 
representative of the LBG population at $z\sim 3$, estimated to be typically 
between $40-300$ Myr \citep[see][]{iwata05}. %To be conservative, the DLA galaxy 
%ages derived for the dwarf irregular bursting model better have to be 
%considered as lower limits (light circles), because the age given in the case 
%of this model corresponds to the time at which the galaxy has started to form 
%stars since its most recent burst. %and hence it does not take into account
%neither the time for the infall of pristine gas nor the existence of possible 
%previous bursts of star formation. 

The DLA galaxies show a large spread in age from a few tens of Myr up to a few 
Gyr. The oldest object is the DLA toward Q\,B2230+02 with an age larger than 
3.2 Gyr. Interestingly, the age of this object, observed at $z_{\rm DLA} = 
1.864$, suggests that galaxies were already forming stars at $z_f\gtrsim 10$. 
The same spread in age is also observed for the emission-selected galaxies at 
similar redshifts. We can note that the age distribution of high-redshift 
galaxies is in agreement with the age limit of the Universe defined by the 
adopted cosmological model at the epoch of observation, but it shows a trend of 
a steeper age decrease with redshift than the age decrease of the Universe. 
This could be due to a selection bias toward the brightest, strongest star 
forming and hence possibly the youngest emission-selected galaxies at high 
redshift that may not be representative of the galaxy population as a whole at 
$z\gtrsim 1.5$. However, the DLA galaxies selected in absorption are 
independent of such a selection bias, and they also show a strong trend toward 
a very young age of a few hundred Myr for galaxies observed at $z > 2$, much 
lower than the age of the Universe at the epoch of observation. This suggests 
relatively low redshifts of formation ($z\sim 3$) for most galaxies observed at 
high redshift.
%for most galaxies observed at $z > 2$.
%JXP -- This is a puzzling and very likely important result.  I think it 
% does say something about z_f (as you have commented out) and/or something
% about mechanisms which 'destroy' DLA.  This is certainly worth a bit
% further discussion (speculation) and possibly a comment in the abstract.

\begin{figure*}[t]
\centering
\includegraphics[width=17cm]{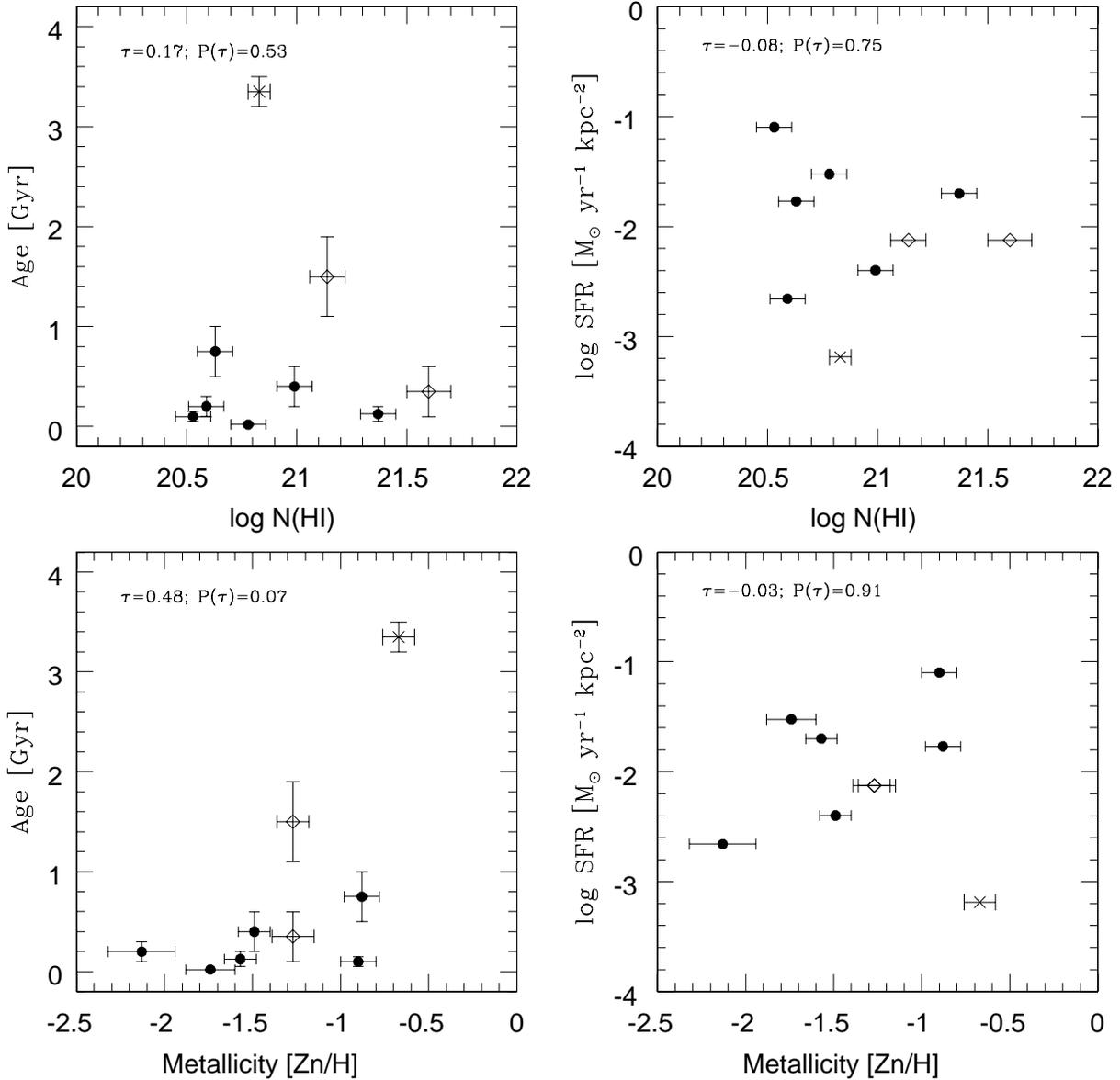}
\caption{Investigation for possible correlations between different physical
properties --~the \ion{H}{i} column density, the metallicity, the age, and the 
star formation rate per unit area~-- derived for the DLA galaxies in our 
sample. The different symbols correspond to the three different star formation
histories identified as best reproducing the abundance patterns of a given DLA: 
spiral outer disk (episodic bursting SFH; diamonds), dwarf irregular bursting 
star formation (single burst SFH; filled circles), and dwarf irregular 
continuous star formation (continuous SFH; crosses). The Kendall rank test was 
performed to analyze these correlations. We give the Kendall rank correlation 
factor, $\tau$, and the probability, P($\tau$), under the null hypothesis of 
zero correlation (values lower than 5\,\% indicate a significant correlation).}
\label{trends}
\end{figure*}

We investigated different possible correlations between the age derived for the 
DLA galaxies in our sample and their \ion{H}{i} column density and metallicity 
properties (see the left panels in Fig.~\ref{trends}). No trend is observed as 
a function of the neutral gas content, while we could have expected a decrease 
of the \ion{H}{i} gas with the galaxy age increase due to the gas consumption 
during the star formation. On the other hand, a trend of metallicity evolution 
with the age of the DLA galaxies seems to be observed. However, due to the low 
number statistics, this correlation mainly relies on one single measurement 
made for the DLA toward Q\,B2230+02 with the highest age and metallicity. But, 
this measurement is very reliable, since this object is one of the rare DLA 
galaxies where star formation history and age were determined in a 
straightforward way; no degeneracy problem was encountered (see 
Sect.~\ref{Q2230}). Such a correlation between age and metallicity may simply 
be explained as the result of the metal gas enrichment of a galaxy with time.
However, if we consider DLA galaxies with the same star formation history, 
for instance only DLAs characterized by a single burst star formation history 
(dwarf irregular bursting models), we can note that the metallicity is, indeed, 
on average dependent on the age of the galaxy, but it also depends on the star 
formation efficiency of the burst. This is illustrated in the right-bottom 
panel of Fig.~\ref{trends} by the tentative correlation between the SFR and 
metallicity observed for DLAs with a single burst SFH.

%
%________________________________________________________________

%
%________________________________________________________________

\subsection{Star formation rate distribution}\label{distr-SFR}

The star formation rates per unit area derived for the nine DLA galaxies in our 
sample are between $-3.2 < \log {\rm SFR} < -1.1$ 
M$_{\odot}$~yr$^{-1}$~kpc$^{-2}$. These values are in very good agreement with
the SFRs obtained by \citet{wolfe03a}, $-3.0 < \log{\rm SFR} < -1.6$ 
M$_{\odot}$~yr$^{-1}$~kpc$^{-2}$, for six DLAs in the redshift interval $z = 
[1.6,2.7]$. \citet{wolfe03a} yet used a very different technique to 
determine the DLA SFRs based on the measure of the cooling rate of these
galaxies assessed by the strength of \ion{C}{ii}$^*$\,$\lambda$1335 absorption. 
The heating rate is then obtained by equating it to the cooling rate, which 
hence constrains the DLA star formation rate, since massive stars forming out 
of neutral gas emit FUV radiation that heats the gas. This shows that the DLA 
galaxies seem really to have moderate to low star formation rates, on average 
similar to the SFR per unit area of the Milky Way Galaxy estimated to 
$4\times 10^{-3}$ M$_{\odot}$~yr$^{-1}$~kpc$^{-2}$ \citep{kennicutt98}. The 
recent work by \citet{wild06} on \ion{Mg}{ii}-selected DLAs at, however, lower 
redshifts $0.4 \leq z_{\rm DLA} \leq 1.3$ also confirms these results. Indeed, 
from the detection of the [\ion{O}{ii}] nebular emission line in 3461 
\ion{Mg}{ii} absorbers, they derived an average SFR per unit area of 
$\log{\rm SFR} \approx -3.0$ M$_{\odot}$~yr$^{-1}$~kpc$^{-2}$.

In Fig.~\ref{age-SFR}\,(b) we compare the DLA star formation rates per unit 
area (circles) with the SFRs obtained for other high-redshift galaxies detected 
in emission: the star-forming galaxies from the Gemini Deep Deep Survey 
\citep[dashed-shaded area;][]{savaglio04}, the LBG MS\,1512$-$cB58 (square), 
and three $\gamma$-ray burst (GRB) host galaxies (diamonds)\footnote{The 
GRB\,00926 at $z=2.04$ \citep{fynbo02}, the GRB\,011211 at $z=2.14$ 
\citep{fynbo03}, and the GRB\,021004 at $z=2.34$ \citep{jakobsson05}.}. The SFR 
of MS\,1512$-$cB58 is representative of the average star formation rate of LBGs 
which have values between $20-370$ M$_{\odot}$~yr$^{-1}$ 
\citep{pettini98,kobulnicky00,iwata05,reddy06}. This clearly shows that the DLA 
galaxies have star formation rates several orders of magnitude (10 to 1000 
times) lower than the SFRs measured in other high-redshift galaxies. Only the
SFRs of GRB hosts\footnote{Called also sometimes GRB-DLA galaxies, since they
are characterized by high \ion{H}{i} column densities as the DLAs and exhibit
damped Ly$\alpha$ profiles in their optical afterglow spectra.} seem to overlap 
with DLAs with the higher SFRs.

Why are the DLA galaxies so different from other known high-redshift 
galaxies\,? There is clearly a strong selection bias toward the more luminous 
and hence toward the stronger star forming galaxies in the sample of 
high-redshift emission-selected galaxies. Hence, in a way the emission-selected 
galaxies are atypical galaxies. The selection of DLA galaxies is, on the 
contrary, largely independent of their luminosity and star formation 
properties. They should thus be representative of the whole galaxy population 
and should account for both the weak and strong star forming galaxies. 
%JXP -- I have rearranged and reworded a bit 
%The privileged selection of 
%weak star forming galaxies with DLAs may hence reflect the indirect effects of 
%three possible selection biases: (i)~the dust obscuration bias against the 
%metal-rich DLA galaxies, i.e. against the dustier and stronger star forming 
%galaxies which would obscure the background quasar more and would therefore be 
%unrepresented in the DLA samples, this bias is still strongly debated 
%\citep[e.g.][]{ellison01,vladilo05}; (ii)~the cross-sectional bias against 
%detecting DLAs in sightlines that pass through the centers of galaxies, which 
%would imply that DLA absorptions mostly select the external, less active in 
%terms of star formation regions of galaxies as supported by theoretical 
%studies of \citet{mathlin01}; and most importantly (iii)~the bias against high
%mass and high SFR galaxies in the early Universe, which would have for 
%consequence that a galaxy sample based on the covering fraction of the sky 
%(cross-section times number) will lead to the detection of lots of low mass and 
%low SFR galaxies.
The predominant identification of weak star-forming galaxies with DLAs may 
hence reflect three indirect effects: 
(i)~high mass and high SFR galaxies are rare in the early Universe, therefore a 
galaxy sample based on the covering fraction of the sky (cross-section times 
number) will generally lead to the detection of lower mass and lower SFR 
galaxies;
(ii)~dust obscuration biases the sample against metal-rich DLA galaxies, i.e. 
against dustier and stronger star forming galaxies which could obscure the 
background quasar
\citep[this bias is still strongly debated, however; see e.g.][]{ellison01,vladilo05}; 
and (iii)~all galaxies have larger cross-section at larger radii, such that DLA 
sightlines tend to avoid the centers of galaxies. This implies that DLA 
absorptions mostly select the external (less active in terms of star formation) 
regions of galaxies as supported by theoretical studies of \citet{mathlin01}. 
Although the latter effects certainly contribute, we expect the dominant effect 
is the former: DLAs sample the broad distribution of galaxies at high redshift, 
a distribution dominated by low masses and low star formation rates.

%JXP -- I am not so sure this following paragraph holds together.  We should
% discuss further.
We finally compare the DLA SFRs obtained from absorption properties with the 
few DLAs for which the emission counterpart was detected and allowed a direct 
SFR estimation (filled triangles in Fig.~\ref{age-SFR}\,(b)). That concerns the 
DLA toward Q\,B2206$-$1958 at $z_{\rm DLA} = 1.919$ 
\citep[e.g.][]{weatherley05}, the DLA SBS\,1543+593 at $z_{\rm DLA} = 0.0096$
toward HS\,1543+5921 \citep{schulte05}, and the DLA toward PKS\,0458$-$02 at 
$z_{\rm DLA} = 2.039$ \citep{moller04}. Interestingly, the DLA SFRs inferred 
from emission show a very large dispersion, much larger than the dispersion 
observed in the SFRs derived from absorption properties; with, in particular, 
the DLA galaxy toward Q\,B2206$-$1958 which reaches a SFR per unit area 
comparable to the median SFR of emission-selected galaxies. The question which 
comes immediately in mind and deserves a paper on itself is: is the SFR of the 
DLA toward Q\,B2206$-$1958 atypical\,? This DLA must in a certain way be 
exceptional, since, despite large observational efforts with at least 20 DLAs 
at high redshift investigated, emission from the associated galaxies is 
generally too weak to be detected \citep[see also][ for a discussion]{wolfe04}. 
If this DLA is not atypical, this would have one major implication: the DLAs 
would hence very likely sample galaxies with a wide range of star formation 
efficiencies and may thus cover a much larger part of the luminosity function 
than expected from the SFR distribution derived from absorption only. This is 
supported by the work of \citet{zwaan05} at $z = 0$. Based on 21~cm column 
density maps of about 400 galaxies, covering all Hubble types and a wide range 
of luminosities, they showed that the distribution of luminosities of the 
galaxies producing DLAs is nearly flat from $M_B = -21$ to $-15$.

We investigated again different possible correlations between this time the 
star formation rates per unit area obtained for the DLA galaxies in our sample
and their \ion{H}{i} column densities and metallicities (see the right panels 
in Fig.~\ref{trends}). No clear correlation is observed. \citet{wolfe03b} 
considered the same quantities and also found no correlation between the SFRs 
and the \ion{H}{i} column densities. They argue that such a correlation should,
however, exist resembling the Kennicutt relation found in nearby disk galaxies
\citep{kennicutt98}. The lack of this correlation may in our work be due to the
fact that with the absorption technique we do not sample the global SFR per 
unit area and the global \ion{H}{i} column density of the galaxy. The values we 
get are sampled over a transverse dimension corresponding to the linear 
dimension of the quasar, i.e. about 1~pc. A dependence of metallicity on the 
SFR is also expected. Wolfe et~al. found marginal evidence for such a 
correlation. We also observe a tentative correlation when considering only the 
DLA galaxies with the same star formation history, namely those characterized 
by a single burst star formation history (dwarf irregular bursting models). 
This is due to the fact that for these objects we have a more direct evidence 
of star formation feedback leading to higher metallicities in presence of 
bursts with higher star formation efficiencies. 

%
%________________________________________________________________

\subsection{Star formation history distribution}\label{distr-SFH}

Three different star formation histories, referred as spiral outer disk 
(episodic bursting SFH), dwarf irregular bursting star formation (single burst 
SFH), and dwarf irregular continuous star formation (continuous SFH), were 
identified as reproducing the DLA abundance patterns (see Sect.~\ref{models}). 
In Figs.~\ref{age-SFR}\,(c) and (d), we show the distributions of the star 
formation histories of the DLA galaxies studied as a function of redshift and 
velocity width of the low-ionization absorption-lines, respectively. The goal 
is to investigate possible trends of specific star formation histories with 
redshift and the mass of the DLA galaxies. We want, however, to underline that 
both the uncertainty existing for most DLAs relative to the degeneracy of their 
best star formation history  and the low number of analyzed objects do not 
allow to draw any statistically significant conclusions. The discussion below 
thus is still preliminary.

The distribution of star formation histories as a function of redshift seems to 
show a trend of finding more galaxies with a star formation history typical of 
dwarf irregulars with a bursting star formation regime toward high redshifts, 
$z>2$. This trend again supports the first effect mentioned in 
Sect.~\ref{distr-SFR} which implies that one has a higher probability to 
encounter younger and less evolved galactic structures with lower masses, 
experiencing their first bursts of star formation, toward higher redshifts. In 
their study of galactic morphologies in the Hubble Deep Field North and South 
as a function of redshift, \citet{conselice05} found a clear decline of the 
number of spirals at $z > 1$. It is thus not surprising to find only a few DLA 
galaxies $-$~two out of nine~$-$ at $z_{\rm DLA} = 1.7-2.5$ having a star 
formation history typical of spiral galaxies. According to the hierarchical 
theory of galaxy formation, these dwarf DLA galaxies at $z\sim 2$ may represent 
the building blocks constituting the spiral galaxies at $z = 0$. This is also 
strongly supported by a straightforward mass argument which shows that the mass 
density in DLAs at $z\sim 2$ is much larger than the mass density in dwarfs 
today \citep{prochaska05}.
%should then primarily evolve into spirals at $z = 0$ after having undergone a 
%series of mergers. 
%JXP -- This is fine, but do keep in mind that these 'dwarf' galaxies at z~2
%  will primarily evolve into spirals at z=0.  The easiest way to see this is
%  by a simple mass argument:  the mass density in DLA at z=2 >> the mass
%  density in dwarfs today (PHW05).

We then examined the distribution of star formation histories as a function of 
the velocity width of the low-ionization lines, likely related to the mass of 
DLA galaxies. Indeed, it was recently suggested that the observed correlation 
between the metallicity and the velocity width of the low-ionization lines in 
DLAs reveals the existence of a mass-metallicity correlation, with the velocity 
width being a mass indicator of these galaxies 
\citep{wolfe98,peroux03,ledoux06}. Here, one would expect a trend of finding 
more galaxies with a star formation history typical of spiral outer disks 
toward high masses. No such a trend is observed, the two DLAs with a spiral 
outer disk star formation history do not have particularly large velocity 
widths relative to the six DLAs with a star formation history typical of dwarf 
irregulars with a bursting star formation. The DLA galaxy with the largest 
velocity width has in fact a star formation history typical of a dwarf 
irregular with a continuous star formation. The star formation history is in 
the case of this DLA toward Q\,B2230+02 very reliable, since it is one of the 
rare DLAs which star formation history determination is not degenerated (see 
Sect.~\ref{Q2230}). Interestingly, this galaxy is also very old with an age 
longer than 3.2 Gyr. As a consequence, the high mass of the galaxy may be 
explained as a requirement to keep the star formation on-going during such a 
long time. 
%On the other hand, the star formation histories typical of dwarf irregulars 
%with a bursting star formation regime have no need of a lot of material in 
%order that a burst happens.

%
%______________________________________________________________

\section{Summary and concluding remarks}\label{conclusions}

This paper presents the star formation properties of a sample of nine DLA 
galaxies at $z_{\rm DLA} = 1.7-2.5$. They were obtained thanks to a new 
technique developed in Paper~I. This technique consists of inferring the star 
formation history and the age of DLA systems from a detailed comparison of 
their intrinsic abundance patterns (free from dust depletion and ionization 
effects) with chemical evolution models. Comprehensive sets of elemental 
abundances are hence required. Our sample of DLAs fulfills this requirement and,
we believe, is a sample representative of the average DLA galaxy population. 
Our results are the first step toward a possible connection between DLA 
galaxies and other galaxy populations identified in deep imaging surveys with 
the aim of obtaining a global picture of high-redshift objects. We obtained the 
following constraints on the DLA galaxies.

(1) Are the DLA abundance patterns consistent with a single class of star
formation history\,? The answer is no. The star formation history was 
unambiguously constrained for three DLAs, and the results clearly show that both 
a continuous and a single burst star formation history typical of local dwarf 
irregular galaxies are demanded. For the six other DLAs, the star formation 
history is degenerate between these two SFHs, but characterized by different 
star formation burst parameters, and the episodic bursting SFH typical of the 
Milky Way type spiral outer disks. Among the burst parameters which were 
investigated, only strong star formation bursts with an efficiency $\nu >> 5$ 
Gyr$^{-1}$ which predict a far too high $\alpha$-enhancement at low and high
metallicities ([Si/Fe] $> +0.5$ at [Fe/H] $< -1$) are ruled out by the observed 
DLA abundance patterns. Otherwise, the full range of dwarf irregular bursting 
models with efficiencies $\nu$ between 0.1 and 4.2 Gyr$^{-1}$ and a relatively 
short burst duration between 0.02 and 0.2 Gyr do satisfy the various DLA 
abundance patterns. Similarly, continuous star formation histories 
characterized by weak star formation efficiencies with $\nu$ values ranging 
from 0.03 to 0.1 Gyr$^{-1}$ do agree with the DLA abundance patterns. However, 
the spiral inner disk models with a continuous star formation history but 
characterized by a 10 times stronger star formation efficiency cannot be ruled 
out, although DLAs intersecting the inner regions of galaxies should be rare. 
Consequently, the abundance patterns show that DLA objects do not represent a 
single class of galaxy population, but sample galaxies with a variety of star 
formation histories. Their common characteristic is that all are weak star 
forming galaxies.

When adopting the star formation histories related to the models which 
reproduce the abundance patterns with the minimal $\chi^2$, we do find only two 
DLA galaxies (each at $z < 2$) in our sample of nine objects with a star 
formation history typical of spiral galaxies. The distribution of the DLA star 
formation histories as a function of redshift shows in fact a trend of finding 
more galaxies with a star formation history typical of dwarf irregulars with a 
bursting star formation toward high redshifts, $z > 2$.  
%JXP -- This sentence was way too long
%Because DLA galaxies are selected solely on their gas cross-section (i.e.\ 
%independently of their luminosities), this trend, if confirmed, indicates, 
%given the DLA covering fraction of the sky, that less evolved galactic 
%structures experiencing their first bursts of star formation and having lower 
%masses as suggested by the distribution of these objects as a function of 
%their line profile velocity widths are more common toward higher redshifts. 
Because DLA galaxies are selected solely on their gas cross-section (i.e.\ 
independently of their luminosities), this trend, if confirmed, indicates, that 
less evolved galactic structures, experiencing their first bursts of star 
formation and having low masses, are more common toward higher redshifts. This 
is further supported by the line profile velocity width, age, and star 
formation rate distributions of the DLA galaxies. 
%JXP -- Is this right?  I would think the degeneracy allows far more than two.
%These DLAs may hence represent the building blocks of today spiral galaxies. 

(2) Are the DLA abundance patterns consistent with a single age\,? The answer 
is no again. The DLA galaxies in our sample show a large spread in age from a 
few tens of Myr up to a few Gyr. The extremities of the spread, 20 Myr and 
$>3.2$ Gyr respectively, correspond to well established ages of the two 
galaxies whose star formation histories are unambiguously constrained. DLAs 
with $\alpha$-enhanced abundance patterns are associated with young objects, 
since the $\alpha$-enhancement is a signature of short-lived massive stars and 
thus of a recent burst of star formation. Solar $\alpha$ over iron-peak element 
ratios, in contrast, are associated with old objects, due to the time necessary 
for the release of iron-peak elements by Type~Ia SNe and the need of an 
inefficient, continuous star formation. The oldest DLA in our sample is 
observed at $z_{\rm DLA} = 1.864$ and has an estimated age larger than 3.2 Gyr. 
It thus nicely indicates that galaxies were already forming at $z_f\gtrsim 10$. 

The general trend which comes out from the DLA age distribution as a function 
of redshift is a much steeper age decrease with redshift than the age decrease 
of the Universe. This is in agreement with the age distribution of 
emission-selected galaxies. A higher proportion of very young galaxies thus 
seem to populate the high redshift Universe at $z > 2$. This suggests 
relatively low redshifts of formation ($z\sim 3$) for most galaxies observed at 
high redshift.

%JXP -- again, I'd prefer we reword the following to minimize the use of bias.
%  It is a fair use of the word but it carries such a negative connotation
%  that I prefer to use 'effects' or something else.
(3) Are the DLA abundance patterns consistent with a single star formation
rate\,? Although the results indicate a spread of two orders of magnitude, this 
is relatively small in comparison with the large range of known star formation 
rates in galaxies. The derived DLA star formation rates per unit area have 
values between $-3.2 < \log {\rm SFR} < -1.1$ M$_{\odot}$~yr$^{-1}$~kpc$^{-2}$, 
in agreement with the SFRs obtained by \citet{wolfe03a} and \citet{wild06} 
using different techniques. In comparison with the SFRs of emission-selected 
galaxies at high redshift, they are 10 to 1000 times lower. Only the SFRs of 
GRB host galaxies seem to overlap with DLAs. 

Why are the DLA galaxies so different from other high-redshift objects\,? 
%JXP -- I'd pose it the other way (why are other high z objects so different
%  from DLA, but that is just me.
The emission-selected galaxy samples are known to be biased toward the more 
luminous and hence the stronger star forming galaxies. But what about DLAs, why 
do they, on the contrary, preferentially select weak star forming galaxies\,? 
This is likely the result of three indirect effects: (i)~high mass and high SFR 
galaxies are rare in the early Universe, therefore a galaxy sample based on gas 
covering-fraction of the sky (i.e.\ the DLA sample), will result in many 
galaxies with lower masses and lower SFRs; (ii)~dust obscuration may bias the 
sample against metal-rich and strong star forming DLA galaxies; and (iii)~the 
cross-section selection implies that DLA sightlines tend to mostly select the 
external, less active regions of galaxies. We expect the dominant effect is the 
former: DLAs sample the broad distribution of galaxies at high redshift, and 
thus this distribution is dominated by young proto-galactic structures with low 
masses and low SFRs. 
%JXP commented out this last bit:  as it is 
%supported by the DLA star formation history and age distributions.

At this stage, to improve our knowledge of the star formation properties of DLA 
galaxies, we really need a large sample of DLAs with detections of their 
emission counterparts. Only the direct conjunction of absorption and emission 
properties of DLA galaxies will bring {\it a major step} in: the comprehension 
of their link with both present-day and other high-redshift galaxies; the 
determination of their contribution to the star formation rate density of the 
Universe; and the proper probe of the galactic chemical evolution.

%(4) Finally, we would like to note that the star formation rates of the few 
%DLAs for which the emission counterpart was detected show a much larger spread 
%than the SFRs derived from absorption properties. In particular, the DLA toward 
%Q2206$-$19 reaches a SFR per unit area comparable to the median SFR of 
%emission-selected galaxies. If this DLA galaxy is not atypical, it may then 
%indicate that the DLAs sample a wide range of the luminosity function and are
%hence representative of the whole galaxy population at high redshifts. However,
%it also suggests the existence of a possible systematic offset between the
%absorption- and emission-measured SFRs, similar to the proposed offset between 
%the metallicities measured in \ion{H}{i} and \ion{H}{ii} regions. This offset
%could typically be the result of the cross-sectional bias already pointed out
%under the point (3).

%
%______________________________________________________________

\begin{acknowledgements}

The authors wish to thank everyone working at ESO/Paranal for the high quality 
of UVES spectra obtained in service mode. We are, in addition, very grateful to 
Cristina Chiappini and Art Wolfe for helpful discussions and advises.
M.D.-Z. and F.C. acknowledge a grant from ESO Office for Science for a visit to 
ESO Garching to work on the analysis of the data. J.X.P. is supported by NSF 
CAREER grant (AST-0548180).

\end{acknowledgements}

%
%______________________________________________________________

\end{document}